\documentclass[prl,aps,floatfix,twocolumn,unsortedaddress]{revtex4}
\usepackage{epsfig}
\begin{document}
%
%
%
%
\title{Bipolar Spintronics: From spin injection to spin-controlled logic}

\author{Igor \v{Z}uti\'{c}}
\affiliation{Department of Physics,
University of Buffalo, SUNY, Buffalo NY 14260}

\author{Jaroslav Fabian}
\affiliation{Institute for Theoretical Physics, University of Regensburg,
93040 Regensburg, Germany}

\author{Steven C. Erwin}
\affiliation{Center for Computational Materials Science,
Naval Research Laboratory, Washington, D.C. 20375}

\begin{abstract}
An impressive success of spintronic applications has been typically realized in 
metal-based structures which utilize magnetoresistive effects for substantial
improvements in the performance of computer hard drives and magnetic
random access memories. Correspondingly, the theoretical understanding 
of spin-polarized transport is usually limited to a metallic regime in 
a linear response, which, while providing a good description for data 
storage and magnetic memory devices, is not sufficient for signal processing 
and digital logic. In contrast, much less is known about possible 
applications of semiconductor-based spintronics and spin-polarized transport 
in related structures which could utilize strong intrinsic 
nonlinearities in current-voltage characteristics to implement spin-based logic.
Here we discuss the challenges for realizing a particular class of structures
in semiconductor spintronics: our proposal for
bipolar spintronic devices in which carriers of both polarities (electrons and
holes) contribute to spin-charge coupling. We formulate the theoretical
framework for bipolar spin-polarized transport, and describe several novel
effects in two- and three-terminal structures which arise from the
interplay between nonequilibrium spin and equilibrium magnetization.
\end{abstract}

\maketitle

\subsection{\label{sec:I} 1. Introduction}

In contrast to well-established applications based on metallic magnetic
multilayers~\cite{Maekawa:2006,Maekawa:2002B,Parkin2003:PIEEE,Prinz1998:S,Ansermet1998:JPCM,%
Hartmann:2000,Hirota:2002,Gregg1997:JMMM,Johnson1994:IEEES,Moodera1999:JMMM,%
Parkin1999:JAP,Tehrani2000:IEEE,Johnson2001:JS,Zutic2004:RMP}, 
much less is known about the prospect for utilizing 
semiconductors in spintronic applications.
Typically, these commercial metal-based applications rely on magnetoresistive
effects and employ two-terminal structures known as the spin-valves in which
a nonmagnetic material is sandwiched between two ferromagnetic electrodes.
The flow of carriers through a spin-valve is determined by the direction of 
their spin (up or down) relative to the magnetization of the device's 
electrodes leading thus to magnetoresistance. Since magnetization in 
ferromagnets persists even when the power is switched off,
these applications have significant advantage of being nonvolatile.
However, for advanced functions, such as signal processing and digital logic, 
two-terminal devices such as those are of limited use. Spin logic will also 
require three-terminal devices and could benefit from incorporating 
semiconductors
with their intrinsic nonlinear current-voltage characteristics, suitable for 
signal amplification. 

\begin{figure}
\centerline{\psfig{file=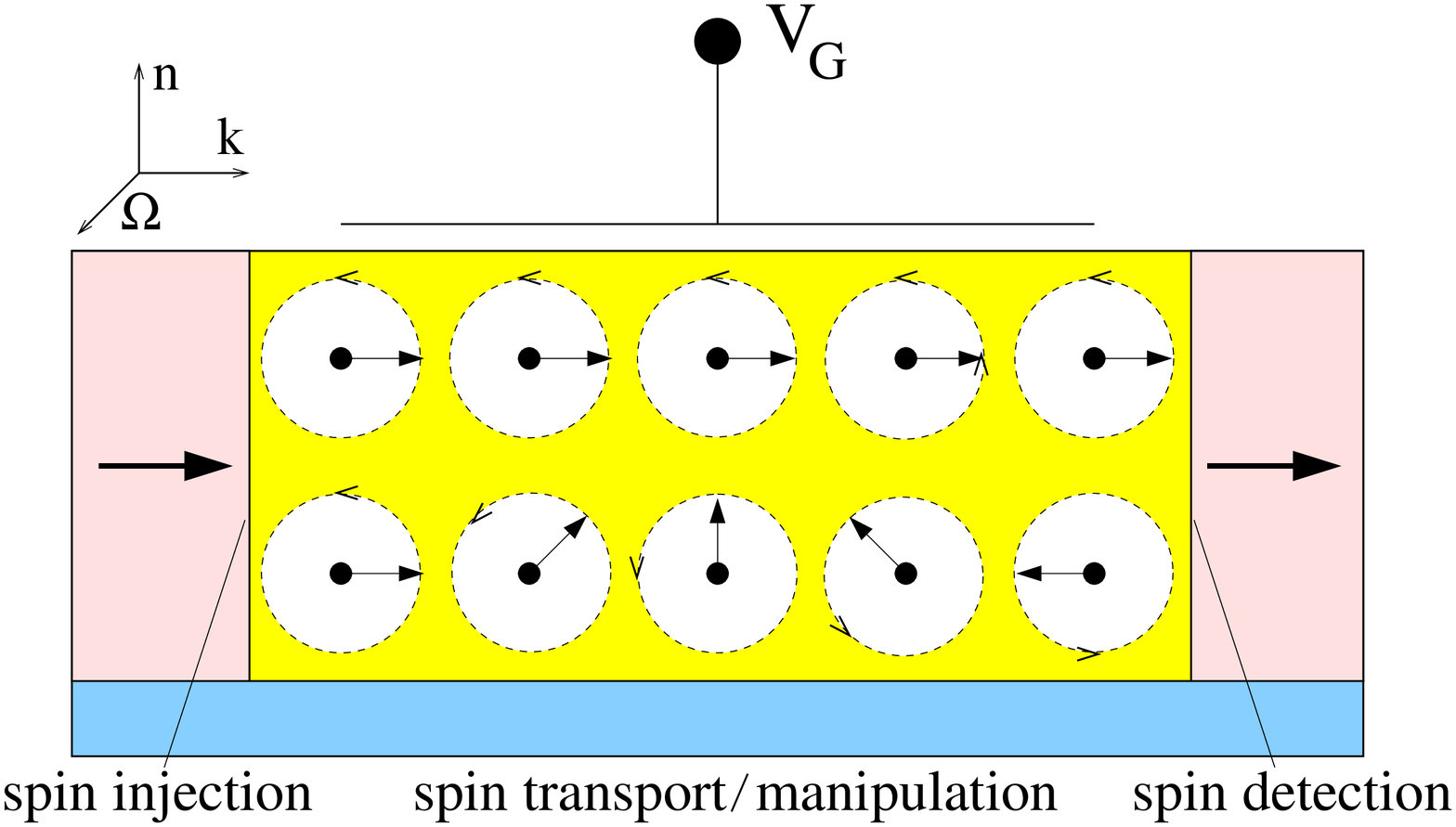,width=1\linewidth,angle=0}}
\vspace{0.5cm}
\caption{Datta-Das spin field effect transistor. The source and drain are
ferromagnetic, while the channel is formed at a heterojunction interface.
The gate modifies the Bychkov-Rashba field {$\bf \Omega$} which is perpendicular
to both the growth direction {\bf n} and electron momenta 
{\bf k}~\cite{Bychkov1984:JPC,Winkler:2003,Zutic2004:RMP}. The electron
either enter in the drain if their spin direction is unchanged (top) or
bounce off if the spin has precessed (bottom), giving ON and OFF states,
respectively. Adapted from~\cite{Zutic2004:RMP}.}
\label{fig:1}
\end{figure}

An early proposal for a semiconductor-based spin-logic device is the Datta-Das
spin field effect transistor (FET)~\cite{Datta1990:APL}, depicted
in Fig.~\ref{fig:1}. While, despite the extensive experimental efforts, 
there remain important challenges for its realization~\cite{Zutic2004:RMP},  
it is helpful to illustrate a generic scheme for a spin logic device with 
basic elements such as spin injection and detection as well as spin transport
and manipulation. 
The spin FET, which can be viewed
as gate-controlled (via spin-orbit coupling) spin-valve, has also spurred many 
related transistor 
schemes~\cite{Wang2002:P,Schliemann2002:P,Matsuyama2002:PRB,Mirales2001:PRB,%
Ciuti2002:APL,Nikonov2005:IEEETN,Sugahara2004:APL,Sugahara2005:JAP,Zutic2004:RMP} . 
However, a similar functionality has
been recently realized in a very different implementation using a carbon nanotube 
(CNT), rather than a semiconductor~\cite{Schapers2001:PRB}, 
as the nonmagnetic material sandwiched between
the ferromagnetic source and drain with tunneling contacts. While a CNT has
a negligible spin-orbit coupling, the tunability of
both the magnitude and the sign of tunneling magnetoresistance in such a CNT 
spin-valve was controlled by gate voltage which changed on or off-resonance
condition~\cite{Sahoo2005:NP,Zutic2005:NP}. Another interesting feature 
of the Datta-Das spin FET 
is that it shows the importance of magnetic heterojunctions as the building 
block for various semiconductor spin-based devices. In this article we will 
review a theory for bipolar spin-polarized transport in magnetic semiconductor
heterojunctions and show possible implications for spin injection and 
spin-controlled logic.
The term {\it bipolar} indicates that carriers of both polarities 
(electrons and holes) are important~\cite{bipolar}. 
In contrast to unipolar devices, 
such as metallic spintronic devices~\cite{Maekawa:2002,Parkin2003:PIEEE},
bipolar devices exhibit large deviations
from local charge neutrality and intrinsic nonlinearities in the
current-voltage characteristics, which are important even at small
applied bias.  

\begin{figure}
\centerline{\psfig{file=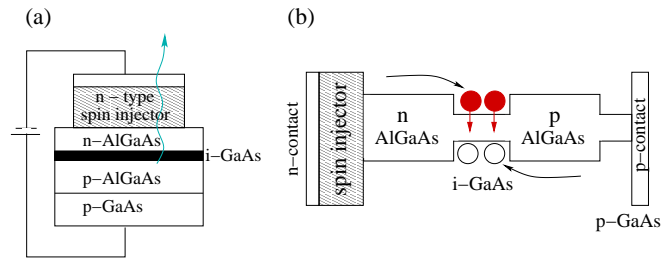,width=1\linewidth,angle=0}}
\vspace{0.5cm}
\caption{Schematic device geometry of a spin LED:
(a) Recombination of spin-polarized electrons injected from
spin injector and  unpolarized holes injected from the
$p$-doped GaAs, in the intrinsic GaAs quantum well,
producing circularly polarized  light;
(b) sketch of the corresponding
band edges and band offsets in the device geometry.
In the quantum well, spin down electrons and unpolarized  holes
are depicted by solid and empty circles, respectively.
Adapted from~\cite{Zutic2004:RMP}.}
\label{fig:2}
\end{figure}

These characteristics, together with the ease of
manipulating the minority charge carriers, enable the design of active
devices that can amplify signals---as well as provide additional
degrees of control not available in charge-based electronics. 
Analogous to the bipolar charge-transport~\cite{Shockley:1950,Sze:1981} 
which is dominated by the influence of the nonequilibrium carrier density,
the nonequilibrium spin density (unequal population of ``spin up'' 
and ``spin down'' carriers) plays an important role in bipolar spintronics.
A spin light emitting diode (LED), depicted in Fig.~\ref{fig:2} can be 
viewed as a prototypical bipolar spintronic device. 
Similar to a an ordinary LED~\cite{Sze:1981},
electrons and holes recombine (in a quantum well or a p-n junction)
and produce electroluminescence. However, in a spin LED, as a consequence
of radiative recombination of spin-polarized carriers, the emitted light
is circularly polarized and could be used to trace back the degree of
polarization of carrier density upon injection into a semiconductor.
While spin LEDs may
not be directly lead to spin logic, they have been widely used as detectors 
for spin polarization, injected optically or electrically into 
a semiconductor~\cite{Fiederling1999:N,Jonker2000:PRB,Young2002:APL,%
Hanbicki2003:P,Jiang2003:PRL,VanRoy2006:MSEB}. 

Another important structure for bipolar spintronics is a 
semiconductor-based magnetic heterojunction and its special cases
such as p-n junctions. In addition to being elements of spin FETs 
and spin LEDs, as we shall show, they are also 
the building blocks for bipolar devices which could enable 
a spin-controlled logic.
Early experimental efforts date back to nearly 40 years ago.
It was shown that a ferromagnetic p-n junction, based on the
ferromagnetic semiconductor CdCr$_2$Se$_4$ doped with Ag acceptors and In
donors, could act as a diode. Photo-voltaic diodes were also fabricated
using (Hg,Mn)Te magnetic semiconductor~\cite{Janik1988:APPA}.
However, a more extensive work on magnetic p-n junction have begun
after the discovery
of (III,Mn)V ferromagnetic semiconductors such as 
(In,Mn)As~\cite{Munekata1989:PRL,Ohno1992:PRL,Munekata1991:JCG},
and  (Ga,Mn)As~\cite{Ohno1996:APL,VanEsch1997:PRB,Hayashi1997:JCG},
reviewed in Refs.~\cite{Ohno1998:S,Dietl2002:SST,Jungwirth2006:P}.
Heavily doped $p$-(Ga,Mn)As/$n$-GaAs junctions
were fabricated~\cite{Ohno2000:ASS,Kohda2001:JJAP,Johnston-Halperin2002:PRB,%
Arata2001:PE,vanDorpe2003:P},
to demonstrate tunneling interband spin injection.
Furthermore, it was shown that the current in $p$-CoMnGe/$n$-Ge magnetic 
heterojunction diodes
can indeed be controlled by magnetic field~\cite{Tsui2003:APL}.

Potentially valuable property for all-semiconductor device designs 
is the external control of Curie temperature ($T_C$). 
Carrier-mediated ferromagnetism in dilute magnetic semiconductors such 
as (In,Mn)As, (Ga,Mn)As, and
MnGe~\cite{Ohno1998:S,Dietl2003:NM,Samarth2003:SSC,Park2002:S,Li2005:APL}
allows for tuning the strength of the ferromagnetic
interactions and, therefore, $T_C$. For example, when the number of
carriers is changed, either by shining
light~\cite{Koshihara1997:PRL,Oiwa2002:PRL} or by applying a gate bias
in a field effect transistor geometry~\cite{Ohno2000:N}, the material
can be switched between the paramagnetic and ferromagnetic states.
These experiments suggest the prospect of nonvolatile multifunctional
devices with tunable optical, electrical, and magnetic properties.
Furthermore, the demonstration of optically or electrically controlled
ferromagnetism provides a method for distinguishing carrier-induced
semiconductor ferromagnetism from ferromagnetism that originates from
metallic magnetic inclusions~\cite{DeBoeck1996:APL}.

An important challenge for potential spin logic applications is 
the demonstration of room temperature operations. 
In all-semiconductor schemes it would be desirable to have ferromagnetic
materials with high $T_C$. Some of the promising developments include
(Ga,Mn)As with $T_C \sim 250$ K~\cite{Nazmul2005:PRL} and 
(Zn,Cr)Te with $T_C \sim 300$ K~\cite{Saito2003:PRL}.
However, there is a wide range of other materials with much higher
predicted and/or reported 
$T_C$~\cite{Dietl2002:SST,Pearton2003:JAP, Erwin2004:NM}
which need to be critically examined and their potential tested in the actual
device structures.

An alternative route to room temperature operation is the
use of hybrid structures that combine metallic
ferromagnets with high $T_C$ and semiconductors.
It is important to note that in such systems tailoring of interfacial
properties can significantly improve magnetoresistive effects or
spin injection efficiency~\cite{Rashba2000:PRB,Smith2001:PRB, Fert2001:PRB,Zega2006:PRL}. For example, a use of MgO (instead 
of Al$_2$O$_3$) as a tunnel barrier between CoFe electrodes
in a magnetic tunnel junction can lead to a dramatic increase in
room temperature tunneling 
magnetoresistance~\cite{Parkin2004:P,Yuasa2004:NM},
confirming previous theoretical 
predictions~\cite{Butler2001:PRB,Mathon2001:PRB}. 
Furthermore, it was demonstrated that employing a CoFe/MgO tunnel 
injector can provide 
robust room temperature spin injection in semiconductors such as 
GaAs~\cite{Jiang2005:PRL,Wang2005:APL} with room temperature
spin polarization of injected electrons exceeding 70 \%~\cite{Salis2005:APL}.

We first formulate drift-diffusion equations for bipolar
spin-polarized transport.
Next we consider spin injection and extraction in magnetic p-n junction 
as well as
an interplay between equilibrium magnetization and the injected
nonequilibrium spin which leads to a strong spin-charge coupling. 
In the last section we review the basics of bipolar junction
transistor and our proposal for its generalization--the magnetic 
bipolar transistor. 

\subsection{\label{sec:II} 2. Bipolar Spin-Polarized Transport}
\subsubsection{\label{sec:IIA} 2.1 Spin-polarized drift-diffusion equations}

Spin-polarized bipolar transport can be thought of as a generalization
of its unipolar counterpart. Specifically, a spin-polarized unipolar
transport, in a metallic regime, can then be obtained as a limiting case by 
setting the
electron-hole recombination rate to zero and considering only
one type of carriers (either electrons or holes).
In the absence of any spin polarization, equations which aim to describe
spin-polarized bipolar transport need to recover a description of
charge transport. A conventional charge transport in 
semiconductors is often accompanied with large deviations from local
charge neutrality (for example, due to materials inhomogeneities, 
interfaces, and surfaces) and Poisson's equation needs to be
explicitly included. If we consider (generally inhomogeneous) doping
with density of $N_a$ ionized acceptors and $N_d$ donors 
we can then write 
\begin{equation}
\nabla \cdot (\epsilon \nabla \phi) = q(n-p+N_a-N_d),
\label{eq:poisson}
\end{equation}
were $n$, $p$ (electron and hole densities) also depend on 
the electrostatic potential $\phi$ and
permittivity $\epsilon$ can be spatially dependent. 
In contrast to the metallic regime,
even equilibrium carrier density can have large spatial variations
which can be routinely tailored by the appropriate choice of the doping
profile [$N_d(x)-N_a(x)$].
Furthermore, charge transport in semiconductors can display strong 
nonlinearities, for example, exponential-like 
current-voltage dependence of a diode~\cite{Sze:1981}.

We briefly recall here a case of a unipolar spin-polarized transport
in a metallic regime. The basic theoretical understanding dates back
to Mott~\cite{Mott1936:PRCa}. He noted that the electrical current
in ferromagnets could be expressed as the sum of two independent
and unequal parts for two different spin projections implying that 
the current is spin polarized. 
We label spin-resolved quantities by
$\lambda=1$ or $\uparrow$ for spin up, $\lambda=-1$ or $\downarrow$
for spin down along the chosen quantization axis.
For a free electron, spin angular momentum
and magnetic moment are in opposite directions, and what
precisely is denoted by ``spin up'' varies in the
literature~\cite{Jonker2003:P}.
Conventionally, in metallic systems~\cite{Gijs1997:AP},
spin up refers to carriers with majority spin. This means that the
spin (angular momentum) of such carriers is
antiparallel to the magnetization.
Some care is needed with the terminology used for semiconductors,
the terms majority and minority there refer to the relative population
of charge carriers (electrons or holes).
Spin-resolved charge current (density)
in a diffusive regime can be expressed as
\begin{equation}
\label{eq:jq}
j_\lambda=\sigma_\lambda \nabla \mu_\lambda,
\end{equation}
where $\sigma_\lambda$ is conductivity and the chemical potential
(sometimes also referred to as the electrochemical potential) is
\begin{equation}
\label{eq:elchem}
\mu_\lambda=(q D_\lambda/\sigma_\lambda) \delta n_\lambda-\phi,
\end{equation}
with $q$ proton charge, $D_\lambda$ diffusion coefficient,
$\delta n_\lambda=n_\lambda-n_{\lambda 0}$ the change of electron density
from the equilibrium value for spin $\lambda$,
and $\phi$ the electric potential. We use a notation in which
a general quantity $X$ is expressed as sum of equilibrium and
nonequilibrium parts $X=X_0+\delta X$. Here we focus on the case of
a collinear magnetization.
More generally, for a noncollinear magnetization, $j_\lambda$ becomes 
a second-rank tensor~\cite{Johnson1988:PRBa,Stiles2002:PRB}.

In the steady state the continuity equation is
\begin{equation}
\label{eq:jcont}
\nabla j_\lambda=
\lambda q \left[\frac{\delta n_\lambda}{\tau_{\lambda-\lambda}}
               -\frac{\delta n_{-\lambda}}{\tau_{-\lambda \lambda}} \right],
\end{equation}
and $\tau_{\lambda\lambda'}$ is the average time for flipping a $\lambda$-spin
to $\lambda'$-spin.
For a degenerate conductor
the Einstein relation
is
\begin{equation}
\sigma_\lambda=q^2 {N}_\lambda D_\lambda,
\label{eq:einstein}
\end{equation}
where $\sigma=\sigma_\uparrow+\sigma_\downarrow$ and
${N}={N}_\uparrow+{N}_\downarrow$
is the density of states. Using a detailed balance
${N}_\uparrow/ \tau_{\uparrow \downarrow}=
{N}_\downarrow/ \tau_{\downarrow \uparrow}$~\cite{Hershfield1997:PRB}
together with Eqs.~(\ref{eq:elchem}) and (\ref{eq:einstein}),
the continuity equation can be expressed 
as~\cite{Rashba2002:EPJ,Takahashi2003:PRB}
\begin{equation}
\label{eq:jq'}
\nabla j_\lambda=\lambda q^2 \frac{{N}_\uparrow {N}_\downarrow}
{{N}_\uparrow+{N}_\downarrow}
\frac{\mu_\lambda-\mu_{-\lambda}}{\tau_s},
\end{equation}
where $\tau_s=\tau_{\uparrow \downarrow} \tau_{\downarrow \uparrow}/
(\tau_{\uparrow \downarrow}+\tau_{\downarrow \uparrow})$
is the spin relaxation time. Equation (\ref{eq:jq'})
implies the conservation of charge
current $j=j_\uparrow+j_\downarrow=const.$, while the spin counterpart,
the difference of the spin-polarized currents
$j_s=j_\uparrow-j_\downarrow$ is position dependent.

Following the work of Mott, a unipolar spin-polarized transport and spin 
injection in the metallic regime is usually described using equivalent resistor 
schemes with two resistors of different magnitudes, one for each spin 
direction, also known as  the ``two-current
model''~\cite{Zutic2004:RMP,Jonker2003:MRS,Jedema2001:N,Valet1993:PRB}.
This approach implies a linear response in which injected spin polarization 
and assumes that there are no interfacial spin-flip processes.
However, the latter assumption, widely used since the first demonstration of spin 
injection in metals~\cite{Johnson1985:PRL}, may need
to be reconsidered~\cite{Rashba2002:EPJ} when analyzing room temperature
spin injection experiments~\cite{Garzon2005:PRL,Godfrey2006:PRL}.

Returning to the case of spin-polarized transport in semiconductors,
we formulate a drift-diffusion model which will generalize the 
considerations of Eq.~(\ref{eq:jq})--(\ref{eq:jq'}) to include
both electrons and holes~\cite{Zutic2001:PRB,Zutic2002:PRL,Fabian2002:PRB}. 
We recall that from  Eqs.~(\ref{eq:jq}) and (\ref{eq:elchem})
spin-resolved current have a drift part (proportional to
electric field i.e., $\propto \nabla \phi$) and a diffusive
part ($\propto \nabla n_\lambda$), which we want to extend 
to also capture the effects of band bending, band offsets, 
various materials inhomogeneities, and the presence of two
type of charge carriers. 
For nondegenerate
doping
levels (Boltzmann statistics) the spin-resolved components are
\begin{equation}
n_\lambda=\frac{N_c}{2} e^{-[E_{c\lambda}-\mu_{n\lambda}]/k_BT}, \quad
p_\lambda=\frac{N_v}{2} e^{-[\mu_{p\lambda}-E_{v\lambda}]/k_BT},
\label{eq:nplambda}
\label{eq:ncv}
\end{equation}
where subscripts $c$ and $v$  label quantities which
pertain to the conduction and valence bands. For example,
$N_{c,v}=2(2 \pi m^*_{c,v} k_B T/h^2)^{3/2}$ are
effective density of states with the corresponding effective masses 
$m^*_{c,v}$ and $k_B$  
is the Boltzmann constant. 
From the total electron density $n=n_\uparrow+n_\downarrow$ 
and the spin density $s=n_\uparrow-n_\downarrow$ we can
define the spin polarization of electron density as
\begin{equation}
P_n=\frac{s}{n}=\frac{n_\uparrow-n_\downarrow}
{n_\uparrow+n_\downarrow}.
\label{eq:pn}
\end{equation}
Such a finite spin polarization does not necessarily require 
ferromagnetic materials or external magnetic fields at all.
For example, circularly polarized
light provides an effective way to generate net spin polarization in
direct-band-gap semiconductors. The angular momentum of the absorbed
light is transferred to the medium; this leads directly to orientation
of the electron orbital momenta and, through spin-orbit interaction,
to polarization of the electron spins~\cite{Meier:1984}. 
In bulk III-V semiconductors,
such as GaAs, optical orientation can lead to 50\% polarization of the
electrons; this can be further enhanced by using quantum structures of
reduced dimensionality, or by applying strain.

\begin{figure}
\centerline{\psfig{file=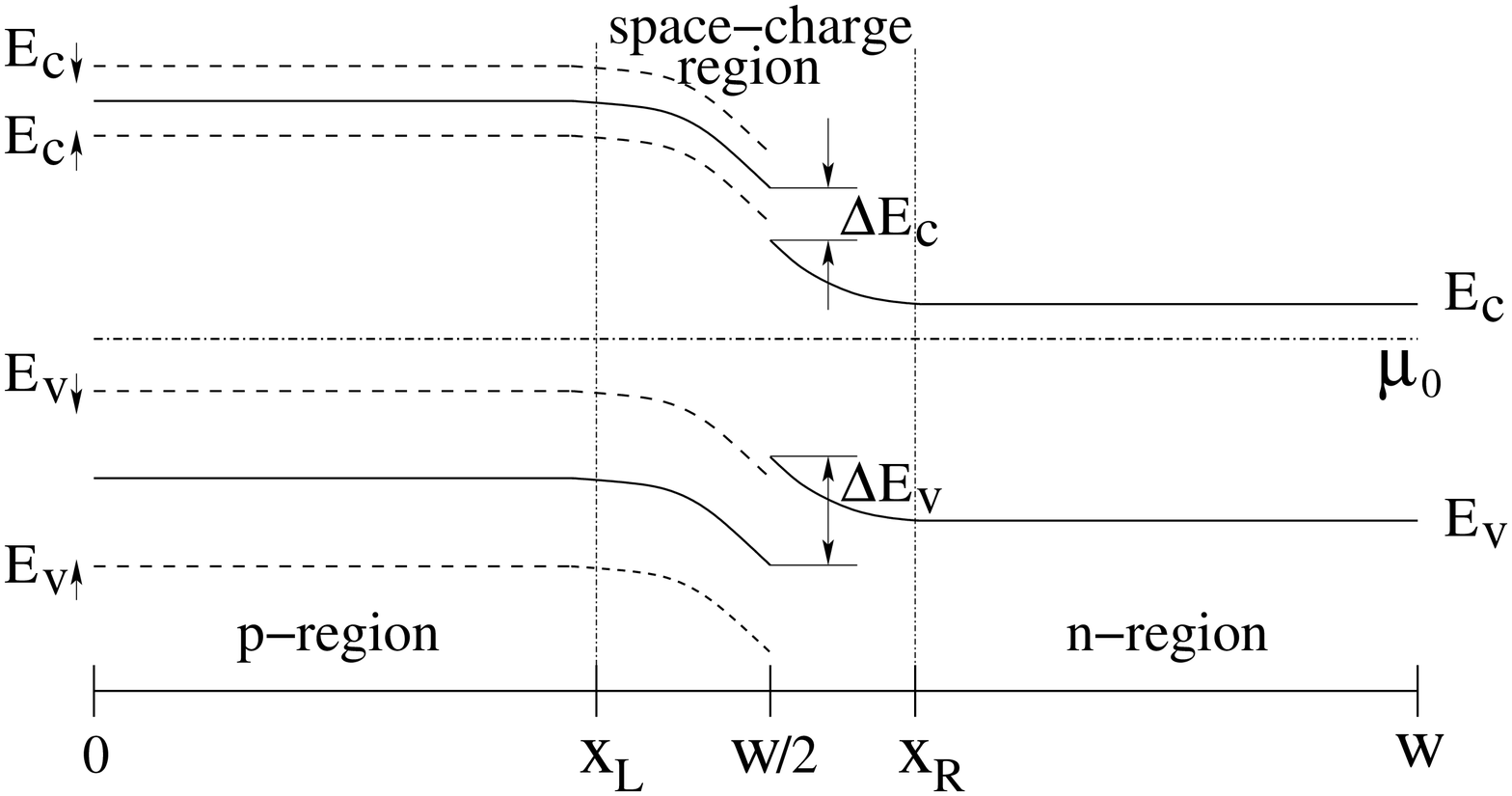,width=1\linewidth,angle=0}}
\vspace{0.5cm}

\caption{Band-energy schemes for a magnetic heterojunction.
In equilibrium chemical potential $\mu_0$ is constant.
Conductance and valence-band edges ($E_c$ and $E_v$) are spin split
in the magnetic p-region, while in the nonmagnetic n-region
there is no spin splitting. Left and right edges of a space-charge 
(depletion) region are denoted by $x_L$ and $x_R$.
For a sharp doping profile, at $x=w/2$, there are generally
discontinuities in conduction and valence bands ($\Delta E_c$
and $\Delta E_v$) and in other quantities, such as,
effective mass, permittivity, and diffusion coefficient. Adapted from Ref.~\cite{Zutic2004:P}}
\label{fig:3}
\end{figure}

We consider a general case where the spin splitting of 
conduction and valence band, expressed as
$2q\zeta_c$ and $2q\zeta_v$, can be spatially 
inhomogeneous~\cite{Zutic2002:PRL}.
Splitting of carrier bands (Zeeman or exchange) can arise due to doping
with magnetic impurities  and/or
applied magnetic field. 
The spin-$\lambda$ conduction band edge (see Fig.~\ref{fig:3})
\begin{equation}
E_{c\lambda}=E_{c0}-q\phi-\lambda q \zeta_c
\label{eq:edge}
\end{equation}
differs from the
corresponding nonmagnetic bulk value $E_{c0}$  due to electrostatic
potential $\phi$ and the spin splitting $\lambda q \zeta_c$.
The discontinuity of the conduction band edge is denoted by
$\Delta E_c$.  In the nonequilibrium state a chemical potential
for $\lambda$-electrons is $\mu_{n \lambda}$ and generally
differs from the corresponding quantity for holes. 
While $\mu_{n \lambda}$ has an analogous role as  
electrochemical potential in Eqs.~(\ref{eq:jq}) and (\ref{eq:elchem}), 
following the conventional semiconductor terminology, we refer to it here 
as the chemical potential, which is also known as the quasi-Fermi level.
An analogous notation holds for the valence band and holes. For example,
in Eq.~(\ref{eq:ncv})  
$p_\lambda$ is the spin-$\lambda$ density of holes with
$E_{v\lambda}=E_{v0}-q\phi-\lambda q \zeta_v$.

By assuming the drift-diffusion dominated transport across a
heterojunction, the spin-resolved charge current densities
can be expressed as~\cite{Zutic2004:P}
\begin{eqnarray}
\label{eq:jn}
{\bf j}_{n\lambda}&=
&\bar{\mu}_{n\lambda} n_{\lambda} \nabla E_{c \lambda}
+qD_{n\lambda} N_c \nabla (n_\lambda/N_c),
\\ \label{eq:jp}
{\bf j}_{p\lambda}&=&
\bar{\mu}_{p\lambda} p_{\lambda} \nabla E_{v \lambda}
-qD_{p\lambda} N_v \nabla (p_\lambda/N_v),
\end{eqnarray}
where $\bar{\mu}$ and $D$ are mobility and diffusion
coefficients (we use symbol $\bar{\mu}$ to distinguish it from
chemical potential $\mu$).
We note that ``drift terms'' have quasi-electric fields  
$\propto \nabla E_{c,v \lambda}$ that are generally
spin-dependent ($\nabla \zeta_{c,v} \neq 0$ is referred to as
a magnetic drift~\cite{Zutic2002:PRL})
and different for conduction and valence bands. 
In contrast to  homojunctions, additional
``diffusive terms'' arise due to the spatial dependence
of $m_{c,v}$, and therefore of $N_{c,v}$.
In nondegenerate semiconductors $\bar{\mu}$ and $D$ are related by Einstein's
relation
\begin{equation}
\bar{\mu}_{n,p \lambda}=q D_{n,p \lambda}/k_BT,
\label{eq:einstein2}
\end{equation}
which differs from the metallic (completely degenerate) case given by
Eq.~(\ref{eq:einstein}). 

With two type of carriers the continuity equations are more complex
than those in metallic systems. After including 
additional terms for recombination of electrons and holes as well as 
photoexcitation of electron-hole pairs, we can write these equations as 
\begin{eqnarray}
-\frac{\partial n_\lambda}{\partial t}+
\nabla\cdot \frac{{\bf j}_{n\lambda}}{q}=&+& r_{\lambda}(n_{\lambda}
p_\lambda -n_{\lambda 0} p_{\lambda 0}) \\ \nonumber
&+& \frac{n_{\lambda}-n_{-{\lambda}}-\lambda \tilde {s}_{n}}{2\tau_{sn}}
-G_{\lambda},
\label{eq:ncont}
\end{eqnarray}
\begin{eqnarray}
+\frac{\partial p_\lambda}{\partial t}+
\nabla\cdot \frac{{\bf j}_{p\lambda}}{q}=&-& r_{\lambda}(n_{\lambda}
p_\lambda -n_{\lambda 0} p_{\lambda 0}) \\ \nonumber
&-& \frac{p_{\lambda}-p_{-{\lambda}}-\lambda \tilde {s}_{p}}{2\tau_{sp}}
+G_{\lambda}.
\label{eq:pcont}
\end{eqnarray}

Generation and recombination of electrons and holes of spin $\lambda$
can be characterized by the rate coefficient $r_\lambda$, the spin relaxation
time for electrons and holes is dented by $\tau_{s n,p}$ and the
photoexcitation rate $G_\lambda$ represents the effects of
electron-hole pair generation and optical orientation.
Spin relaxation equilibrates carrier spin while preserving 
nonequilibrium carrier density and for nondegenerate semiconductors
$\tilde{s}_n=n P_{n0}$, where from Eq.~(\ref{eq:ncv}) an equilibrium 
polarization of electron density is 
\begin{equation}
P_{n0}=\tanh(q \zeta_c/k_BT),
\label{eq:pn0}
\end{equation}
and an analogous expression holds for holes and $\tilde{s}_p$.

The system of drift-diffusion equations (Poisson and continuity 
equations) can be self-consistently solved 
numerically~\cite{Zutic2001:PRB,Zutic2001:APL,Zutic2002:PRL} and
under simplifying assumptions (similar to the case of charge transport)
analytically~\cite{Fabian2002:PRB,Zutic2003:APL,Zutic2004:P}.
Heterojunctions, such as one sketched in Fig.~\ref{fig:3},
can be thought of as building blocks of bipolar spintronics.
To obtain a self-consistent solution in such a geometry, 
only the boundary conditions at $x=0$ and $x=w$ need to be specified.
On the other hand, for an analytical solution we also need to specify
the matching conditions at $x_L$ and $x_R$, the two edges of the space 
charge region (or depletion region), in which there is a large deviation 
from the local charge neutrality, accompanied by a band bending and strong
built-in electric field. 

We illustrate how the matching conditions for spin and carrier
density can be applied within the small-bias 
or low-injection approximation, widely 
used to obtain analytical results for charge 
transport~\cite{Sze:1981,Ashcroft:1976}. 
In this case nonequilibrium carrier densities are small compared to the 
density of majority carriers in the corresponding semiconductor region.
For materials such as GaAs a small bias approximation gives a
good agreement with the full self-consistent solution up to 
approximately 1 V~\cite{Zutic2001:APL,Zutic2002:PRL}. 
To simplify our notation, we consider a model where only electrons are
spin polarized ($p_\uparrow=p_\downarrow=p/2$),
while it is straightforward to also include
spin-polarized holes~\cite{Zutic2004:P,Fabian2002:PRB}. 
Outside the depletion charge region materials parameters 
(such as, $N_a$, $N_d$, $N_c$, $N_v$, $\bar{\mu}$, and $D$) are taken to
be constant. The voltage drop is confined to the depletion 
region which is highly resistive and depleted from carriers.
In thermal equilibrium $(\mu_{n \lambda} =\mu_{p \lambda}=\mu_0)$ 
the built-in voltage $V_{bi}$ can be simply evaluated 
from Eq.~(\ref{eq:ncv}) as 
\begin{equation}
V_{bi}=\phi_{0R}-\phi_{0L},
\label{eq:builtin} 
\end{equation}
while the applied bias $V$ (taken to be 
positive for forward bias) can be expressed as
\begin{equation}
V=-(\delta \phi_R-\delta \phi_L),
\label{eq:applied} 
\end{equation}
implying that the total junction potential between $x=0$ and $x=w$
is $V-V_{bi}$. 
For a heterojunction sketched in Fig.~\ref{fig:3} 
the width of a depletion (space-charge) region is 
\begin{equation}
x_R-x_L \propto \sqrt{V_{bi}-V},
\label{eq:width}
\end{equation}
where the built-in voltage is
$q V_{bi}=-\Delta E_c+k_B T \ln (n_{0R} N_{cR}/n_{0L} N_{cL})$.
Outside of the depletion region  
the system of drift-diffusion equations reduces usually to only 
diffusion equations for spin density and the density of minority carriers,
while the density of majority carriers are simply given by the density of
donors and acceptors~\cite{Zutic2002:PRL,Fabian2002:PRB}. 
These diffusion equations contain spin and charge diffusion
lengths 
\begin{equation}
L=\sqrt{D \tau},
\label{eq:L}
\end{equation}
in which $L$ would provide a characteristic lengthscale for the spatial decay
of nonequilibrium spin or charge by substituting for $D$ the appropriate
(electron or hole) diffusion coefficient and for $\tau$ (spin or charge)
the characteristic timescale. 
However, there are situations, due to additional effects of spin-orbit
coupling or simultaneous spin polarization of electrons and holes
in magnetic semiconductors, in which diffusion equations become
more complicated and Eq.~\ref{eq:L} 
needs to be generalized~\cite{Tse2005:PRB,Zutic2004:P}.  
For several decades the techniques of 
optical orientation have been used to directly measure the characteristic
timescale for the decay of nonequilibrium electron spin~\cite{Meier:1984}
reaching up to 30 ns~\cite{Weisbuch1977:T}.
More recent optical measurements have shown at low temperatures even 
longer spin lifetime in GaAs ($> 40$ ns) ~\cite{Dzhioev1997:PSS} and  
($> 100$ ns)~\cite{Kikkawa1998:PRL,Dzhioev2001:JETPL} 
which could reach $\sim 1$ ns at room temperature. Spin-orbit coupling
in the valence band typically leads to much faster spin relaxation of
holes than electrons (spin lifetimes are 3-4 orders of magnitude shorter
in GaAs at room temperature~\cite{Hilton2002:PRL}) further supporting
our approximation of spin-unpolarized holes.
The related issues of spin relaxation and spin dephasing in GaAs
have been extensively reviewed in Ref.~\onlinecite{Zutic2004:RMP}.

From Eq.~(\ref{eq:ncv}) we rewrite electron density by separating
various quantities into equilibrium and nonequilibrium parts as
\begin{equation}
n_\lambda=n_{\lambda 0} 
\exp[(q \delta \phi+\delta \mu_{n \lambda})/k_BT],
\label{eq:nlambda}
\end{equation}
and electron  carrier and  spin density (for simplicity we omit
subscript "$n$"  when writing $s=n_\uparrow-n_\downarrow$)
can be expressed as~\cite{Fabian2002:PRB} 
\begin{equation}
n=e^{(\delta \phi+\delta \mu_+)/k_BT}
  \left[ n_0\cosh \left (\frac{q \mu_-}{k_BT} \right )
+s_0 \sinh \left(\frac{q \mu_-}{k_BT} \right) \right ], 
\label{eq:ntot}
\end{equation}
\begin{equation}
s=e^{(\delta \phi+\delta \mu_+)/k_BT}
  \left[ n_0\sinh \left( \frac{q \mu_-}{k_BT} \right)
+s_0 \cosh \left( \frac{q \mu_-}{k_BT} \right ) \right ], 
\label{eq:stot}
\end{equation}
where $\mu_\pm \equiv (\mu_{n\uparrow} \pm \mu_{n\downarrow})/2$,
and the polarization of electron density is
\begin{equation}
P_n=\frac{\tanh(q \mu_-/k_BT)+P_{n0}}
{1+ P_{n0} \tanh(q \mu_-/k_BT)}.
\label{eq:ptot}
\end{equation}

If we assume that the spin-resolved chemical potentials are constant
for $x_L  \leq x  \leq x_R$ [which means that the depletion region
is sufficiently narrow so that the spin relaxation and
carrier recombination can be neglected there] it follows, 
from Eq.~(\ref{eq:ptot}) and $\tanh(q\mu_-/k_BT) \equiv const.$, 
that  
\begin{equation}
P_{n}^L=\frac{P_{n0}^L[1-(P_{n0}^R)^2]+
\delta P_{n}^R(1-P_{n0}^LP_{n0}^R)}
{1-(P_{n0}^R)^2+\delta P_{n}^R(P_{n0}^L-P_{n0}^R)},
\label{eq:aL}
\end{equation}
where $L$ (left) and $R$ (right) label the edges of the
space-charge (depletion) region of a p-n junction. Correspondingly,
$\delta P_n^R$ represents  the nonequilibrium
electron  polarization, evaluated at $R$, arising from a
spin source. 
For a homogeneous
equilibrium magnetization ($P_{n0}^L=P_{n0}^R$),
$\delta P_{n}^L=\delta P_{n}^R$;
the nonequilibrium spin polarization is the same across the depletion region.
Equation (\ref{eq:aL}) demonstrates that only {\it nonequilibrium} spin,
already present in the bulk region, can be transferred through the depletion
region at small biases~\cite{Zutic2001:PRB,Zutic2002:PRL,Fabian2002:PRB}.

Our assumption of constant spin-resolved chemical potentials
is a generalization of a conventional models for charge transport
in which both $\mu_n$ and $\mu_p$ are assumed to be constant across
the depletion region~\cite{Ashcroft:1976}. From Eqs.~(\ref{eq:applied}),
(\ref{eq:ntot}), and (\ref{eq:stot}) we can   
obtain  minority carrier and spin densities at $x=x_L$
\begin{equation}
n_L=n_{0L}e^{qV/k_BT}\left[1+
\delta P^R_n \frac{P^L_{n0}-P^R_{n0}}{1-(P_{n0}^R)^2} \right],
\label{eq:nL}
\end{equation}
\begin{equation}
s_L=s_{0L}e^{qV/k_BT}\left[1+
\frac{\delta P^R_n}{P^L_{n0}} 
\frac{1-P^L_{n0}P^R_{n0}}{1-(P_{n0}^R)^2} \right],
\label{eq:sL}
\end{equation}
which in the absence of nonequilibrium spin ($\delta P^R_n=0$)
reduce to the well-known Shockley relation for the minority
carrier density at the depletion region~\cite{Shockley:1950}
\begin{equation}
n_L=n_{0L}e^{qV/k_BT},
\label{eq:shock}
\end{equation}
and an analogous formula holds for spin 
\begin{equation}
s_L=s_{0L}e^{qV/k_BT}.
\label{eq:shocks}
\end{equation}

\subsubsection{\label{sec:II2} 2.2 Magnetic p-n junctions}
Even in nonmagnetic p-n junctions the presence of nonequilibrium
spin (created electrically or optically) can have interesting implications.
By the term nonmagnetic we imply the limit of vanishing equilibrium magnetization
or, equivalently, vanishing spin polarization since $\zeta_{c,v}=0$. 
We have predicted that the nonequilibrium spin polarization is bias dependent.
By analogy with junction capacitance, this effect could be called spin-capacitance
as the amount of accumulated spin changes with applied bias. In contrast to the
usual monotonic spatial decay of spin density in the nonmagnetic 
metal~\cite{Zutic2004:RMP}, away from the point of spin injection, 
in inhomogeneously doped semiconductors
(such as p-n junctions) the spatial profile can be qualitatively
different. Spin density can even increase inside the nonmagnetic
region, away from the point of spin injection, which we refer to 
as (spatial) spin density amplification~\cite{Zutic2001:PRB,Zutic2006:IBM}. 
Illumination of p-n junction by circularly polarized light
can lead to spin electromotive force (EMF) to generate spin-polarized currents
even at no applied bias and to provide an open-circuit voltage.
In addition to our proposal for a p-n junction-based spin-polarized
solar battery~\cite{Zutic2001:APL}, there is a range of other structures which
could be used as a source of spin 
EMF~\cite{Ganichev2003:JPCM,Long2002:APL,Malshukov2003:PRB} 
or which could display the
effects of spin capacitance~\cite{Datta2005:APL} 
and spatial spin 
amplification~\cite{Ganichev2003:JPCM,Long2002:APL,Pershin2003:PRL}.   

To take an additional advantage of manipulating spin degrees of freedom 
which could lead to strong spin-charge coupling and potential
applications~\cite{Zutic2004:RMP}, it is useful to consider
semiconductor structures with equilibrium magnetization. Such magnetization or,
equivalently, equilibrium spin polarization arising from carrier spin-subband
splitting (see Fig.~\ref{fig:3}) is readily realized in applied magnetic field.
Effective $g$-factors can be much larger than for free electrons, either due to
magnetic impurities~\cite{Furdyna1988:JAP} 
($|g| \approx 500$ at $T < 1$ K~\cite{Dietl:1994}, 
in $n$-doped (In,Mn)As $|g| > 100$ at 30 K~\cite{Zudov2002:PRB}) or due to
strong spin-orbit coupling in narrow-band gap semiconductors (in InSb
$|g| \approx 50$ even at room temperature). 
In the absence of magnetic field, equilibrium magnetization and spin 
splitting can be realized using ferromagnetic semiconductors.
Magnetic impurities and/or an application of an inhomogeneous 
magnetic field
could be used to obtain a desirable, spatially inhomogeneous, spin splitting.
Inhomogeneous spin splitting can also occur in domain walls,
discussed, for example, in Ref.~\cite{Deutsch2004:JAP}. 
By solving a system of drift-diffusion and Poisson equations, 
one can show that an inhomogeneous spin splitting leads to deviations 
from local charge neutrality~\cite{Fabian2002:PRB}.

\begin{figure}
\centerline{\psfig{file=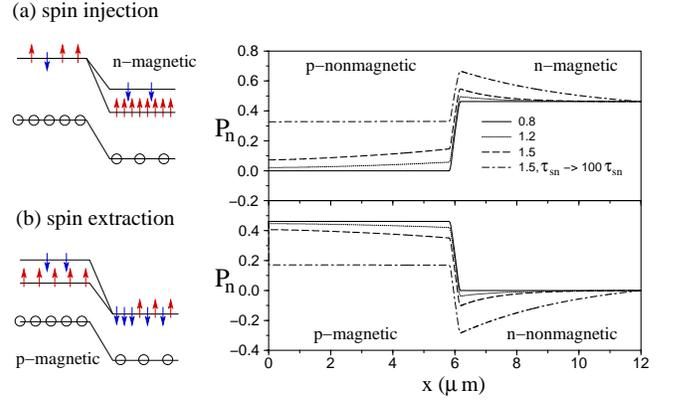,width=1\linewidth,angle=0}}
\vspace{0.5cm}
\caption{Spin injection and extraction in magnetic p-n junctions.
(a) Band-energy diagram with spin-polarized electrons (arrows) and 
unpolarized holes (circles) for spin injection. 
Spin polarization profiles for different 
applied bias (labeled by the numbers in volts) show that the 
spin polarization injected in the nonmagnetic region increases with bias 
and spin relaxation time $\tau_{sn}$. 
For $V \lesssim 0.8$ V there in only a negligible spin 
injection in the p-region while there is a sizeable equilibrium spin
polarization in the magnetic n-region, determined by the spin splitting of 
$q \zeta =0.5 k_BT$.
(b) Band-energy diagram for spin extraction. 
Electrons from nonmagnetic n-region preferentially populate a lower 
spin level in the magnetic p-region which leads to spin extraction. 
In contrast to spin injection, spin polarization profiles show
opposite signs in the nonmagnetic and magnetic regions. 
Adapted from~\cite{Zutic2002:PRL}.}
\label{fig:4}
\end{figure}

We discuss several properties of magnetic p-n junctions 
which rely on the interplay of the carrier spin-subband splitting
and the nonequilibrium spin induced, for example,
by optical or electrical means. 
We also focus here on a diffusive regime while a magnetic diode in a 
ballistic regime was recently discussed in 
Ref.~\cite{Schmeltzer2003:PRB}.
For simplicity, we look at a particular case where the band-offsets 
(see, Fig.~\ref{fig:3}) 
are negligible and the spin polarization of holes can be neglected and both in
the notation for the carrier spin-splitting $2 q \zeta$ and
for the spin density $s$ we can omit index $n$. 
From Eqs.~(\ref{eq:ncv}) and (\ref{eq:edge})
we can rewrite product of equilibrium densities as
\begin{equation}
n_0 p_0=n_i^2\cosh(q\zeta/k_BT),
\label{eq:mass}
\end{equation}
where $n_i$ is the intrinsic (nonmagnetic) carrier 
density~\cite{Ashcroft:1976} and we notice 
that the density of minority carriers in the p-region will
depend on the spin splitting 
$n_{0}(\zeta)=n_{0}(\zeta=0)\cosh(q\zeta/k_B T)$.
In Fig.~\ref{fig:4} we illustrate bipolar spin-polarized transport across a 
magnetic p-n junction under applied forward bias.
Calculations are performed using a self-consistent solution of system of 
drift-diffusion and Poisson equations.
The parameters taken
for $w=12$ $\mu$m long junctions are based on GaAs-like material doped with
$N_a=3\times10^{15}$ cm$^{-3}$ acceptors to the left and 
$N_d=5\times10^{15}$ cm$^{-3}$ donors to the right. Diffusion coefficients
are $D_n^L=10D_p^R=103.6$ cm$^2$s$^{-1}$, the intrinsic carrier density
is $n_i=1.8\times10^6$ cm$^{-3}$, the permittivity is $\epsilon=13.1$,
recombination rate coefficient is 
$r_\uparrow=r_\downarrow=(2/3)\times 10^{-5}$ cm$^3$s$^{-1}$, and the
spin relaxation time is $\tau_{sn}=0.2$ ns. The minority diffusion
lengths are~\cite{Zutic2001:PRB,Zutic2001:APL} $L_n=1$ $\mu$m, 
$L_p=0.25$ $\mu$m, and the electron spin diffusion length in the
n-(p-)region is $L_{sn}=1.4$ $\mu$m ($L_{sp}=0.8$ $\mu$m).

We first ask whether spin can be injected and extracted into/from
the nonmagnetic region. At small bias ($V<V_{bi}\approx 1.1$ V) there
is no spin injection or extraction. As the bias increases, the injection
and extraction become large and are further enhanced with $\tau_{sn}$. 
The reason why there is no spin injection or extraction at small bias
is that although there are exponentially more spin up than spin 
down electrons (recall the Boltzmann statistics) in the magnetic side, the
barrier for crossing the space-charge region is exponentially larger for
spin up than for spin down electrons (see Fig.~\ref{fig:4}). Those
two exponential effects cancel out, leaving no net spin current flowing
through the space-charge region. We could examine these arguments with
analytical findings from Eq.~(\ref{eq:aL}) also valid for low-bias regime.
Considering spin injection in the nonmagnetic n-region ($P_{n0}^R=0$)
we see that in the absence of nonequilibrium spin polarization 
at $x=x_R$, $\delta P_n^R=0$, 
there is indeed no spin injection: $P_n^L=P_{n0}^L$.
This is in contrast to arguments which suggest that an inefficient spin 
injection arises from resistance (conductivity) 
mismatch~\cite{Schmidt2002:SST}. Here
we note that even with a good conductivity match 
and highly-polarized spin 
injector the spin injection can still be completely negligible. 
Furthermore, the efficiency of spin injection depends strongly on the 
applied bias rather than on the relative conductivities of the two regions.

At large bias self-consistent numerical results become indispensable,
showing that spin injection/extraction is possible as a result of building 
up a nonequilibrium spin at the space-charge region. However, some of
these trends, including our prediction for spin extraction,
can already be seen analytically in the low bias regime,
Using the previous assumption that $\mu_{n\uparrow,\downarrow}$
are constant for $x_L \leq x \leq x_R$ and by solving diffusion equations
for $x < x_L$ and $x > x_R$ we can obtain~\cite{Fabian2002:PRB}
\begin{equation}
\delta s_R= -\gamma_3 s_{0L} e^{qV/k_BT},
\label{eq:extract}
\end{equation}
where 
\begin{equation}
\gamma_3=\left( \frac{D_n^L}{D_n^R} \right ) 
\left( \frac{L_{sn}}{L_n} \right )
\frac{\tanh[(w-x_R)/L_{sn}]}{\tanh(x_L/L_n)},
\label{eq:gamma3}
\end{equation}
implies that a contribution to the nonequilibrium spin in the 
n-region will have the
opposite sign to that of the equilibrium spin in the p-region.
Similar spin extraction was recently observed experimentally in 
MnAs/GaAs junctions~\cite{Stephens2004:PRL} and related 
theoretical implications due to tunneling from 
nonmagnetic semiconductors into metallic 
ferromagnets were considered~\cite{Bratkovsky2004:JAP}.
Furthermore, it was suggested that a combination of spin injection and 
spin extraction could lead to completely spin polarized carriers
in semiconductor nanostructrures~\cite{Osipov2005:APL,Petukhov2006:P}. 

\begin{figure}
\centerline{\psfig{file=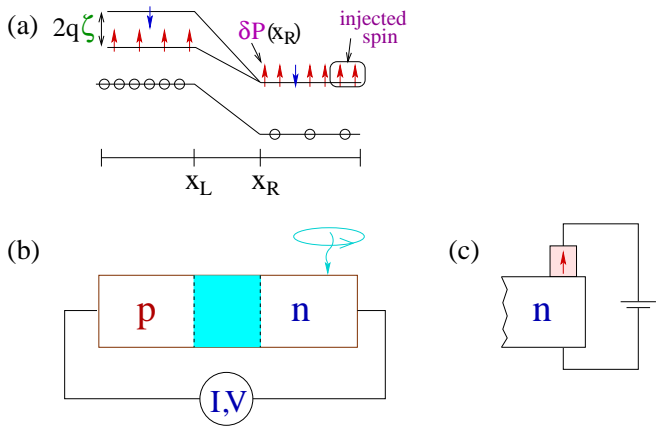,width=1\linewidth,angle=0}}
\vspace{0.4truecm}
\caption{Scheme of a magnetic p-n junction which could lead
to the spin-voltaic effect and giant magnetoresistance. (a) Band-energy diagram. 
The spin splitting $2q\zeta$, the nonequilibrium spin polarization at the 
depletion region edge $\delta P_n(x_R)$, and the region where the spin is 
injected are depicted.
(b) Circuit geometry corresponding to panel (a). Using circularly
polarized light
(photoexcited electron-hole pairs absorb the angular momentum
carried by incident photons), nonequilibrium spin is injected transversely in 
the nonmagnetic n-region and the circuit loop for  $I-V$ characteristics 
is indicated.
Panel (c) indicates an alternative scheme to electrically inject spin into the
n-region. Adapted from~\cite{Zutic2003:APL}.
}
\label{fig:nano1}
\end{figure}

We next consider a simple scheme of a magnetic p-n junction, 
depicted in Fig.~\ref{fig:nano1}, in which there is an external source of 
nonequilibrium spin, induced optically or electrically. As we discuss below,
an interplay between the equilibrium spin polarization and the nonequilibrium spin
source leads to the spin-voltaic effect (a spin analog of the photo-voltaic
effect) and to giant magnetoresistance.

Similar to the theory of charge transport in nonmagnetic 
junctions~\cite{Shockley:1950} the total charge current can be 
expressed as the sum of minority carrier currents at the depletion edges 
$j=j_{nL}+j_{pR}$ with
\begin{equation}
j_{nL} \propto \delta n_L, \quad j_{pR} \propto \delta p_R,
\label{eq:jLR}
\end{equation}
where $\delta n_L$ is given by Eq.~(\ref{eq:nL}) with 
$P^R_{n0}=0$,  $\delta p_R=p_0[\exp(qV/k_BT)-1]$, and $V$ is the
applied bias (positive for forward bias).
Eq.~(\ref{eq:mass}) implies that in the regime of large spin splitting, 
$q \zeta > k_BT$, the density of minority electrons changes exponentially 
with $B$ ($\propto \zeta$) and can give rise to exponentially large 
magnetoresistance~\cite{Zutic2002:PRL}.

The interplay between the $P_{n0}$ [recall Eq.~(\ref{eq:pn0})]
in the p-region, and the nonequilibrium
spin source of polarization $\delta P_n$
in the n-region, at the edge of the depletion region,
modifies the $I-V$  characteristics of the diodes.
To illustrate the $I-V$ characteristics of magnetic p-n junction, 
consider the small-bias 
limit 
[recall Eqs.~(\ref{eq:builtin})-(\ref{eq:shocks})]
in the configuration of Fig.~\ref{fig:nano1}. 
The electron contribution to the total electric current can
be expressed from Eqs.~(\ref{eq:nL}) and (\ref{eq:jLR}) 
as~\cite{Zutic2002:PRL,Fabian2002:PRB}
\begin{equation}
j_{nL} \sim n_{0}(\zeta)
\left [e^{qV/k_B T} \left (1+\delta P_n P_{n0}\right ) -1 \right ].
\label{eq:md}
\end{equation}
Equation (\ref{eq:md})
generalizes the
Silsbee-Johnson spin-charge coupling~\cite{Silsbee1980:BMR,Johnson1985:PRL},
originally proposed for ferromagnet/paramagnet metal interfaces, to
the case of magnetic p-n junctions.  The advantage
of the spin-charge coupling in p-n junctions, as opposed to metals or
degenerate systems, is the nonlinear voltage dependence
of the nonequilibrium carrier and spin 
densities~\cite{Zutic2002:PRL,Fabian2002:PRB},
allowing for
the exponential enhancement of the effect with increasing $V$. 
Equation (\ref{eq:md}) can be understood qualitatively
from Fig.~\ref{fig:nano1}.
In equilibrium, $\delta P_n=0$
and $V=0$, no current flows through the
depletion region, as the electron currents from both sides of the junction
balance out. The balance is disturbed either by applying bias or
by selectively populating different spin states,
making the flow of one spin species greater than that of the other.
In the latter case,
the effective barriers for crossing
of electrons from the $n$ to the $p$ side is different for spin
up and down electrons (see Fig.~\ref{fig:nano1}).
Current can flow even at $V=0$ when $\delta P_n\ne 0$.
This is an example of the spin-voltaic effect, 
in which  nonequilibrium
spin causes an EMF~\cite{Zutic2002:PRL,Zutic2003:P} and
the direction of the zero-bias current
is controlled by the relative sign of
$P_{n0}$ and $\delta P_n$. 
We emphasize here that the spin-voltaic effect results in a build up of 
electrical voltage due to the proximity of the equilibrium and
nonequilibrium spin. This effect is distinct from the so-called spin Hall
effect(s) which results in a build up of a spin imbalance (different
chemical potential for spin up and down), but no electric field, due to
transport in a spin-orbit field~\cite{Tse2005:PRB}.

A straightforward method for detecting the spin-voltaic effect 
follows from the symmetry
properties of the different contributions to the charge current under
magnetization reversal.
By reversing the equilibrium spin polarization using
a modest external magnetic field ($P_{n0} \rightarrow -P_{n0}$)
a part of $j_{nL}$, odd under such reversal, can be identified
as the spin-voltaic current $j_{sv}$~\cite{Zutic2002:PRL}.
Measurements of  $j(V,P_{n0})-j(V,-P_{n0})
=2 j_{sv}(V,P_{n0})$ would then provide: (1) cancellation
of contributions to the charge current that are not related
to the injected spin; (2) a choice of
$V$ to facilitate
a sufficiently large $j_{sv}$ for accurate detection.
Unlike the spin LEDs, this approach does not rely on direct band-gap material
and injected spin could be detected even in silicon~\cite{Zutic2004:P}.
Magnetic semiconductors approximately lattice matched with Si could be used for
spin injection and detection [(Ga,Mn)As was already grown on 
Si~\cite{Zhao2002:JCG}]. For 
example,
the Mn-doped chalcopyrite
ZnGeP$_2$ (mismatch $< 2\%$)~\cite{Ishida2003:PRL} 
has been reported to be ferromagnetic at
room temperature. Another Mn-doped chalcopyrite, ZnSiP$_2$, was
recently predicted~\cite{Erwin2004:NM} to be ferromagnetic, as well as
highly spin polarized and closely lattice-matched with Si (mismatch
$<1\%$). Mn doping of the chalcopyrite alloy ZnGe$_{1-x}$Si$_x$P$_2$
would likely lead to an exact lattice match, since the lattice
constant of Si is between those of closely matched ZnSiP$_2$ and
ZnGeP$_2$.

Additionally, from $j_{sv}(V)$ one could also determine a spin relaxation
time by all-electrical means~\cite{Zutic2003:APL}.
A particular assumption of a magnetic homojunction is not essential. One
could also generalize this analysis to heterojunctions which would include
(see Fig.~\ref{fig:3}) band offsets and spin splitting in both conduction
and valence bands. 

Several experimental efforts have recently observed the spin-voltaic effect
in semiconductor heterojunctions.
One of the approaches used {\it p-n} (In,Ga)As/(Al,Ga)As 
heterojunction~\cite{Kondo2006:JJAP} in the applied magnetic field.
Circularly polarized light was used to inject nonequilibrium spin in (Al,Ga)As while 
an applied magnetic field created equilibrium spin splitting in (In,Ga)As [a particular 
Al-composition in (Al,Ga)As-region can produce nearly zero $g$-factor].
An interesting implication of this device is that it operates as 
a spin-photodiode~\cite{Kondo2006:JJAP}.
By converting circular polarization directly into an electrical signal it is a
counterpart of a spin LED which converts electrical signal into emission of
circularly polarized light. In another approach both the spin injection and 
the detection were realized electrically. Iron was used as a spin injector
into n-doped GaAs, while the spin splitting in p-doped (Ga,Mn)As enabled
spin detection~\cite{Chen2006:P}.  

We will revisit the implications of spin-voltaic effect in three-terminal
structures, discussed in the section on magnetic bipolar transistors.

\begin{figure}
\centerline{\psfig{file=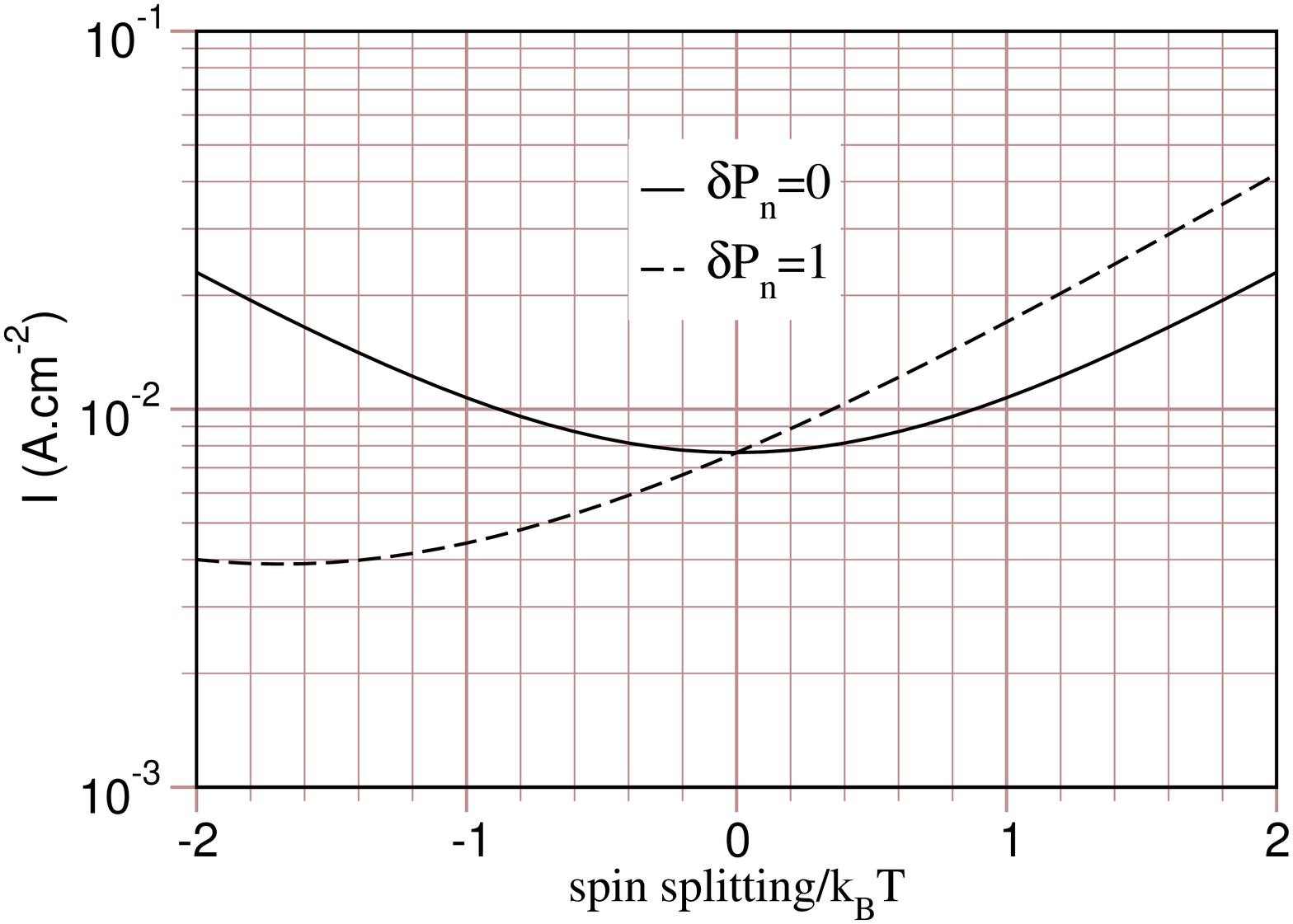,width=0.9\linewidth,angle=0}}
\caption{Giant magnetoresistance-like effect in magnetic p-n junctions.
Current/spin-splitting characteristics ($I$-$\zeta$)
are calculated self-consistently
at $V$=0.8 V
for the geometry from  Fig.~\ref{fig:nano1}.
Spin splitting $2q\zeta$ on the $p$-side is normalized to $k_B T$.
The solid curve, 
symmetric in $\zeta$,
corresponds to a switched-off spin source. 
With spin source on (the extreme case of 100\% spin
polarization injected into the
n-region
is shown), the current is a strongly asymmetric
function of $\zeta$, displaying large GMR, shown by the dashed curve.
Materials parameters of GaAs, used in Fig.~\ref{fig:4},  were applied.
Adapted from~\cite{Zutic2002:PRL}.}
\label{fig:gmrd}
\end{figure}

In addition to the spin-voltaic effect, the spin-charge coupling in magnetic p-n 
junctions can produce a
giant magnetoresistance (GMR)-like effect, which
follows from Eq.~(\ref{eq:md})~\cite{Zutic2002:PRL}. The
current depends strongly on the relative orientation of the
nonequilibrium spin and the equilibrium magnetization.
Figure \ref{fig:gmrd} plots $j$,
which also includes the contribution from holes, as a function of
$2q\zeta/k_B T$ for
both the unpolarized, $\delta P_n=0$, and fully polarized, $\delta P_n=1$,
n-region.
In the first case $j$ is
a symmetric function of $\zeta$, increasing exponentially with increasing
$\zeta$ due to the increase in the
equilibrium minority carrier density $n_0(\zeta)$.
In unipolar systems, where transport is due to the majority carriers,
such a modulation of the current is not likely, as the majority carrier
density is fixed by the density of dopants.
A realization of exponential magnetoresistance was recently
demonstrated in a very different materials system of 
manganite-titanate heterojunctions~\cite{Nakagawa2004:P} in
which an applied magnetic field affected the width 
of a depletion layer.

If $\delta P_n \ne 0$, the current will depend on the sign of
$P_{n0}\cdot \delta
P_n$.  For parallel nonequilibrium (in the n-region)
and equilibrium spins (in the p-region), most electrons cross
the depletion region through the
lower barrier (see Fig.~\ref{fig:nano1}), increasing the current.
In the opposite case of antiparallel relative orientation,
electrons experience a larger barrier and the current is inhibited.
This is demonstrated in Fig.~\ref{fig:gmrd}
by the strong asymmetry in $j$. The corresponding GMR ratio, the
difference between  $j$ for parallel and antiparallel orientations,
can also be calculated analytically from Eq.~(\ref{eq:md}) as
$2|\delta P_n P_{n0}|/(1-|\delta P_n P_{n0}|)$~\cite{Fabian2002:PRB}.

\subsection{3. Spin Transistors}

Thus far we have mostly considered two-terminal spintronic devices in which
we were concerned with spin injection and spin-voltaic phenomena.
However, the greatest strength of the semiconductor spintronics
should lie in the possibility to fabricate three-terminal structures
that would allow current gain. Two goals can be set: First, to
extend the functionalities of the existing transistors by adding
spin control, and, second, to improve the performance of the current
technology in terms of speed, power consumption, or sensitivity.
Whether or not these goals will be reached depends much on the
progress in fabrication and materials development, as well as on our
understanding of the physics of spin-charge coupling in
semiconductor heterojunctions.

We will briefly discuss a few proposed and existing spin transistor designs
(some of them are further reviewed in Ref.~\cite{Zutic2004:RMP}),
before we analyze in more detail our proposal for magnetic bipolar
transistors. The canonical example of a spin transistor is that of
Datta and Das~\cite{Datta1990:APL}, depicted in Fig.~\ref{fig:1}. 
The
gating does not involve charge build up so the transistor could be
faster and less power consuming than conventional field effect
transistors. In addition, the magnetic configurations of the
electrodes can be useful for storing information. 
However, despite many 
experimental efforts, the Datta-Das transistor has not been
realized. There are inherent difficulties in its semiconductor-based design, 
the most important being the spin injection to the quasi one dimensional
conduction channel (which, as noted in Sec.~1, could be avoided by using
carbon nanotubes). An interesting alternative to the spin
field-effect transistor, a spin MOSFET, has been a more conventional
structure 
employing ferromagnetic source and drain and using the spin-valve effect to
control 
the current~\cite{Sugahara2004:APL, Sugahara2005:JAP}.
Since the proposed structure includes silicon substrate it could be 
potentially useful for silicon spintronics.
An important prerequisite for spin MOSFET would be a demonstration of efficient 
spin injection in Si~\cite{Zutic2004:P} and there are recent efforts 
to fabricate suitable ferromagnet/Si contacts~\cite{Min2006:NM,Zutic2006:NM}. 
Another effort to incorporate
silicon in spintronics devices relies on spin diffusion transistor
with silicon base~\cite{Dennis2005:JMMM}. 
The emitter and collector contacts are
ferromagnetic metal-insulator-semiconductor junctions and early
encouraging results show both magnetoresistive effects and current
gain greater than unity.

There have been other transistor designs using metallic layers in the
structure. They go under the name of hot electron spin transistors
or spin-valve transistors~\cite{Zutic2004:RMP,Jansen2003:JPD}. 
For example, one realization (which is also
called the magnetic tunneling transistor) uses a combination of a
ferromagnetic tunnel junction and a Schottky barrier 
collector~\cite{Mizushima1997:IEEETM, Yamauchi1998:PRB, Sato2001:APL,%
vanDijken2002:APL, vanDijken2003:APLa, vanDijken2003:PRL}. The
tunnel junction plays the role of the emitter-base junction,
supplying hot spin-polarized electrons into the magnetic base. The
base-collector junction is a Schottky barrier. The hot electrons
from the base can overcome the barrier only if their energy is
higher than the barrier height. Since the hot electrons lose their
energy depending on the spin, the magnetic junction can effectively
control the collector current: for parallel magnetizations of the
junction the current is large, while for antiparallel it is small,
since the spin up, say, hot electrons in a spin down base
equilibrate more efficiently than in a spin up base. This is the
physics behind the high magnetocurrent rations (reaching thousands
of percents) in these transistors. The disadvantage of these 
hybrid
metal/semiconductor transistors is the absence of gain (the word
transistor here points to the three-terminal geometry rather than to
the ability to amplify currents). Nevertheless the hybrid designs
have been successful in achieving the large magnetocurrent ratios
and spin injection into semiconductors. 
A direct connection of such
structures with bipolar transport, discussed in this manuscript, was 
recently realized in the spin-valve structures which contain a
nonmagnetic p-n junction which can have useful effects 
as the energy barrier~\cite{Huang2004:APL,Huang2005:JAP}.

Motivated by the potential ease of the integration of magnetic
semiconductors with conventional devices we have proposed what we
call magnetic bipolar transistors (MBT), in which one or more
regions (emitter, base, and collector) are formed by a magnetic or
ferromagnetic semiconductor~\cite{Fabian2002:P}. We have shown that
such structures can exhibit giant magnetoamplification, a
significant control of electrical properties by magnetic field, as
well as spin injection all the way from the emitter to the
collector. The magnetic bipolar transistor was later discussed in
terms of spin currents in Ref.~\onlinecite{Flatte2003:APL} and in
terms of magnetoamplification in Ref.~\onlinecite{Lebedeva2003:JAP}.
A bipolar transistor-like scheme has been recently presented in 
Ref.~\cite{Dery2006:PRB}: the semiconductor spin-diffusive channel is
topped with three ferromagnetic electrodes, enabling the
amplification of magnetoresistance.

\subsection{4. Magnetic Bipolar Transistors}

\begin{figure}
\centerline{\psfig{file=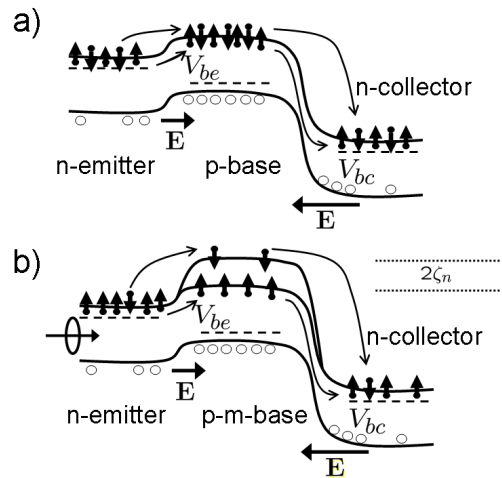,width=1\linewidth,angle=0}}
\caption{a) Scheme of a conventional bipolar junction transistor.
The conduction and valence band profiles are shown. The emitter and
the collector are n-type, the base is p-type. The electrical field
directions in the two depletion layers are indicated. The
base-emitter voltage is $V_{be}$, while the base-collector voltage
is $V_{bc}$. In the active forward regime, shown in the scheme, the
electrons from the emitter go through the forward biased
base-emitter junction ($V_{be}
> 0$) into the base, where they diffuse towards the collector while
also partially recombining with the holes in the base. Those which
reach the collector will be swept by the huge built-in field into
the collector, participating in the collector current. (b) The
corresponding magnetic bipolar transistor with the spin splitting
$2q \zeta_n$ of the conduction band in the base (``p-m'' refers
to magnetic p-doping). In addition, a source
spin can be injected into the emitter, enabling spin-charge effects
such as giant magnetoamplification. Holes are assumed unpolarized.}
\label{fig:transistor1}
\end{figure}

\subsubsection{4.1 Connection to Bipolar Junctions Transistors}
We next  describe the
operation principles of the magnetic bipolar transistor and
their conventional (nonmagnetic) counterparts.
Consider an npn transistor, as in Fig.~\ref{fig:transistor1}a. The
transistor consists of three regions: emitter, base, and collector.
There are two p-n junctions in series: base-emitter and
base-collector. Depending on the polarity of the bias across the
junctions the transistor exhibits different functionalities. The
current gain in bipolar junction transistors appear only in the
active forward and active reverse regimes. In both regimes one
junction is forward, the other reversed biased. The only distinction
between the two regime comes from the asymmetry of the actual
device. In the active forward bias the base-emitter junction is
forward biased, while the base-collector junction is reverse biased.
In the active reverse regime the bias polarities as switched.
Typically the emitter is more heavily doped than the collector,
which results in a much greater current gain for the active forward
regime due to the increased electron injection efficiency from the
emitter (see below).

There are two more possibilities for the transistor operation. In
the saturation mode both of the junctions are forward biased. The
transistor exhibits no gain, but this configuration is used in logic
operations to represent the ON state. Similarly, the configuration
in which both junctions are reverse biased, also called cutoff, is
used to represent the OFF state.

We will see that those configurations have a much richer structure
in magnetic bipolar transistors, shown schematically in
Fig.~\ref{fig:transistor1}b.
 In fact, there exists an additional
configuration, in which both junctions are unbiased but spin-charge
coupled (through the spin-voltaic effect), 
which can be used as a spin switch. All the possible
functionalities are summarized in Table~\ref{tab:transistor}.

\begin{table}
\begin{tabular}{|c|c|c|c|c|}
\hline
mode & $V_{be}$ & $V_{bc}$  & BJT & MBT  \\
\hline \hline
forward active & F & R & amplification & MA, GMA \\
\hline
reverse active & R & F & amplification & MA, GMA \\
\hline
saturation & F & F & ON & ON, GMA, SPSW \\
\hline
cutoff & R & R & OFF & OFF \\
\hline
spin-voltaic & 0 & 0 & OFF & SPSW \\
\hline
\end{tabular}
\caption{Operational modes of bipolar junction transistors (BJT) and
magnetic bipolar transistors (MBT). Forward (F) and reverse (R) bias
means positive and negative voltage, respectively. Symbols MA and
GMA stand for magnetoamplification and giant magnetoamplification,
while ON and OFF are modes of small and large resistance,
respectively; SPSW stands for spin switch. Adapted from
Ref.~\onlinecite{Fabian2005:APL}.} 
\label{tab:transistor}
\end{table}

We have seen in the Sec.~2.2 on magnetic diodes that the
spin-charge coupling (or more specifically, the spin-voltaic effect)
across magnetic p-n junctions can either intensify or inhibit
carrier injection. The electron current in a magnetic junction
was given by Eq.~(\ref{eq:md}). This spin-charge
coupling induces significant changes in the operation of the
magnetic bipolar transistor in terms of what we have named
magnetoamplification and giant magnetoamplification 
(GMA)~\cite{Fabian2002:P,Fabian2004:APL,Fabian2004:PRB,Fabian2004:APP}. 

Let us first see what is the mechanism behind the current gain in
conventional bipolar transistors. The amplification factor $\beta$
is customarily written as~\cite{Tiwari:1992}
\begin{equation}
\beta = \frac{1}{\alpha_T' + \gamma'}.
\end{equation}
Here
\begin{eqnarray}
\alpha_T' & = &  \frac{1-\alpha_T}{\alpha_T}, \\
\gamma' & = & \frac{1-\gamma}{\gamma},
\end{eqnarray}
and $\alpha_T$ and $\gamma_T$ are the base transport factor and the
emitter injection efficiency, respectively. We can then call
$\alpha_T'$ and $\gamma'$ the base transport {\it in}efficiency,
which is due to the electron-hole recombination in the base, and
emitter injection {\it in}efficiency, respectively.

The base transport inefficiency is given by the expression
\begin{equation}
\alpha_T' = \cosh\left (\frac{w_b}{L_{nb}} \right ) -1,
\end{equation}
which is valid for both conventional and magnetic transistors. Here
$w_b$ is the width of the base and $L_{nb}$ is the electron
diffusion length. Typically the width of the base, $w_b$, is much
smaller than the electron diffusion length in the base, $L_{nb}$, in
which case $\alpha_T' \approx (w_b/L_{nb})^2/2$. Typically $L_{nb}$
is about a micron, so the base transport factor does not play a
significant role in modern bipolar transistors with a narrow base.
Another possibility how to reduce the base factor is to employ
graded semiconductors with built-in electric fields that help the
diffusing electrons from the emitter to reach the collector faster
to inhibit electron-hole recombination.

Unlike the base inefficiency, the emitter inefficiency depends
rather strongly on the spin properties of the magnetic transistor.
Let us write
\begin{equation}
\gamma' = \gamma_0'\eta,
\end{equation}
where $\gamma_0'$ is the emitter inefficiency for conventional
spin-unpolarized transistors,  while the spin-charge coupling is
contained in the factor $\eta$. We can then write, for the practical
limit of $\alpha_T' \ll \gamma'$, that
\begin{equation}
\beta \approx \eta /\gamma_0' \approx \eta \beta_0.
\end{equation}
We will be concerned only with $\eta$; $\beta_0$ is the conventional
transistor current gain. For us it is just a numerical factor of the
order of a hundred.

\subsubsection{4.2 Magnetoamplification}

In the absence of an external spin injection, all the spin
properties of magnetic bipolar transistors are determined by the
equilibrium magnetization of the transistor regions. Similarly to
magnetic diodes, this equilibrium magnetization influences
electrical properties as well, due to the influence on the
equilibrium minority carrier density. Consider a magnetic base, for
example, assuming its spin polarization to be $P_{0b}$. The
equilibrium electron density there is
\begin{equation}
n_{0b} = \frac{n_i^2}{N_{ab}} \frac{1}{\sqrt{1-P_{0b}^2}},
\end{equation}
where $n_i$ is the intrinsic density of the underlying
semiconductor material and $N_{ab}$ is the acceptor doping. Since
the current across p-n junctions, in our case the emitter-base
junction, is linearly dependent on $n_{0b}$, the electron emitter
current and the emitter efficiency will be linearly proportional to
$n_{0b}$ as well. The spin-charge factor $\eta$ then stands out as
\begin{equation}
\eta = \frac{1}{\sqrt{1-P_{0b}^2}}.
\end{equation}
This finally gives for the gain of the transistor
\begin{equation}
\beta =\frac{\beta_0}{\sqrt{1-P_{0b}^2}}.
\end{equation}
It turns out that the gain can be controlled also by the emitter
spin polarization~\cite{Lebedeva2003:JAP}, but is unaffected by the
possible equilibrium collector spin~\cite{Fabian2004:PRB}. The
control of the current gain by the equilibrium spin polarization has
been termed magnetoamplification~\cite{Fabian2002:PRB}.

\subsubsection{4.3 Giant magnetoamplification}

Giant magnetoamplification is a direct consequence of the
spin-charge coupling across a p-n junction. Consider a magnetic
bipolar transistor with a magnetic base having equilibrium spin
polarization $P_{0b}$. The emitter and the collector are
nonmagnetic. Suppose we can excite nonequilibrium spin in the
emitter, giving it a nonequilibrium spin polarization $\delta P_e$.
As a result of the proximity of the equilibrium and nonequilibrium
spin there appears an EMF across the junction and the modification
of the electron injection efficiency. This spin-charge coupling is
reflected in the $\eta$:
\begin{equation}
\eta = \frac{1+\delta P_e P_{0b}}{\sqrt{1-P_{0b}^2}}.
\end{equation}
The current gain factor becomes
\begin{equation}
\beta = \beta_0\frac{1+\delta P_e P_{0b}}{\sqrt{1-P_{0b}^2}}.
\end{equation}

The spin dependence of the current gain due to the spin-charge
coupling has been termed giant magnetoamplification~\cite{Fabian2004:PRB}, 
due to the potentially giant relative
difference of amplification for parallel ($\delta P_e P_{0b} > 0$)
and antiparallel ($\delta P_e P_{0b}<0$) configurations. The
corresponding giant magnetoamplification coefficient is
\begin{equation}
{\rm GMA} = \frac{\beta_{p} - \beta_{ap}}{\beta_{ap}},
\end{equation}
where the subscripts $p$ and $ap$ represent the parallel and
antiparallel configurations, respectively.
For the above case of the magnetic base we obtain [recall
also Sec.~2.2 and Eq.~(\ref{eq:md})]
\begin{equation}
{\rm GMA} = \frac{2|\delta P_e P_{0b}|}{1-|\delta P_e P_{0b}|}.
\end{equation}
If the equilibrium and nonequilibrium spin polarizations were about
50\%, the corresponding GMA factor would be about 67\%.

\subsubsection{4.4 Spin injection}

We have seen in the previous section that spin injection from a
magnetic n-region to a nonmagnetic p-region of a p-n junction is not
possible at small biases due to balancing thermodynamics of the
equilibrium spin polarization and the spin-dependent thermal
activation. Only nonequilibrium spin can be injected, that is, the
(nonequilibrium) spin has to first accumulate in the magnetic
region.

Although the magnetic bipolar transistor comprises two magnetic p-n
junctions in series, spin injection is possible even at low biases
in the forward active regime.  
The reason for the possibility of spin
injection is that minority electrons injected from the emitter to
the magnetic base accumulate in the base. There is thus
nonequilibrium electron density with a gradient sufficient to drive
the electrons by diffusion to the collector. The electron spins
equilibrate in the base to the equilibrium spin polarization. We
thus have nonequilibrium spin {\it density} in the base, with the
equilibrium spin {\it polarization}. It is this nonequilibrium spin
density that drives the spin injection to the collector: as the
spin-polarized electrons move towards the base-collector depletion
layer, they are swept by the built-in electric field of the layer to
the collector. In effect, we have a minority electron spin pumping,
similar to what happens in spin-polarized p-n junctions or solar
cells~\cite{Zutic2001:PRB}. The resulting spin polarization in the
collector, in the limit of a narrow base (the width smaller than the
spin diffusion length), is~\cite{Fabian2004:PRB}
\begin{equation} \label{eq:dPc1}
\delta P_c \approx  P_{0b} \frac{L_{sc}}{w_b}
\frac{e^{qV_{be}/k_BT}}{N_{dc}},
\end{equation}
where $L_{sc}$ is the spin diffusion length in the collector and
$N_{dc}$ is the collector donor density. For a spin polarization of
$P_{0b} \approx 0.5$ one can achieve $\delta P_c$ as large as 0.1,
mainly due to the large ratio of the base width and the spin
diffusion length in the collector.

Much less surprising is the spin injection possibility for a source
spin from the emitter to the collector. Suppose we induce a spin
polarization $\delta P_e$ in the emitter. In the active forward
regime the spin is injected to the collector through the following
sequence of steps: First, the source spin diffuses towards the
base-emitter junction. If the spin diffusion length is larger than
the length the spin travels from the injection point to the
base-emitter depletion layer, the spin will attenuate only a little.
Then the spin is transferred to the base through the depletion
region. If we for now take a nonmagnetic base, the spin polarization
in the base will be roughly $\delta P_e$. The nonequilibrium spin
polarizations at the two sides of a p-n junction are the same, as
follows from the generalized Shockley theory of spin-polarized p-n
junctions~\cite{Fabian2002:PRB}. As a result of the nonequilibrium
spin polarization, there appears nonequilibrium spin {\it density}
in the base. This density will be injected to the collector as a
result of the minority electron spin pumping, similarly to the case
of the no source spin case above. We have derived the following
formula for the injected spin polarization in the collector:
\begin{equation}
\delta P_c \approx  \delta P_e \frac{L_{sc}}{w_b}
\frac{n_{0b}e^{qV_{be}/k_B T}}{N_{dc}}.
\end{equation}
Comparing with Eq.~(\ref{eq:dPc1}), we see that the role of the source
spin polarization in the emitter is similar to the role of the
equilibrium spin in the base. Both can be efficiently injected
through to the collector.

In summary, magnetic bipolar transistors offer new functionalities
to conventional semiconductor electronics. The most exciting is the
possibility of large magnetoamplification effects. 
In all other
aspects the magnetic bipolar transistors will have a similar performance to
their conventional counterparts, since they are based on the same physical
principles governing electronic transport.

\subsection{5. Conclusions}

We have reviewed here several phenomena associated with bipolar spin-polarized
transport in semiconductors. Our findings for two-terminal structures, such
as magnetic heterojunctions, can also be applied to more complicated multi-terminal
geometries.
We show that the interplay of magnetic region with equilibrium spin polarization
and injected nonequilibrium spin leads to the spin-voltaic effect in a heterojunction. 
This theoretical prediction,
a spin-analog of the photo-voltaic effect, was
also recently confirmed experimentally. 
The direction of the charge current, which can flow even at no applied bias,
can be switched by reversal of the equilibrium magnetization or by reversal
of the polarization of the injected spin.
In three-terminal magnetic bipolar transistors the spin-voltaic effect 
implies that one could effectively control the gain or current
amplification in such devices.  We predict the possibility for giant
magnetoamplification which could be viewed as a generalization of the
spin-valve effect to semiconductor structures with strong intrinsic
nonlinearities suitable for spin-controlled logic.

\subsection{Acknowledgements}
We thank S. Das Sarma, M. Fuhrer, B. T. Jonker,  
H. Munekata, S. S. P. Parkin, A. Petrou, A. Petukhov, and E. I. Rashba 
for useful discussions.
This work was supported by the US ONR, NSF-ECCS CAREER, DARPA,
NRC-NRL, and SFB 689.  
This work was performed in part at SUNY's Buffalo Center for Computational
Research and used the resources of the Center for Computational Sciences
and  the Center for Nanophase Materials Sciences
at Oak Ridge National Laboratory, which is supported by the Office of
Science of the U.S. Department of Energy under
Contract No. DE-AC05-00OR22725.
\bibliographystyle{apsrev}
\bibliography{referencesJPCM}

\begin{thebibliography}{149}
\expandafter\ifx\csname natexlab\endcsname\relax\def\natexlab#1{#1}\fi
\expandafter\ifx\csname bibnamefont\endcsname\relax
  \def\bibnamefont#1{#1}\fi
\expandafter\ifx\csname bibfnamefont\endcsname\relax
  \def\bibfnamefont#1{#1}\fi
\expandafter\ifx\csname citenamefont\endcsname\relax
  \def\citenamefont#1{#1}\fi
\expandafter\ifx\csname url\endcsname\relax
  \def\url#1{\texttt{#1}}\fi
\expandafter\ifx\csname urlprefix\endcsname\relax\def\urlprefix{URL }\fi
\providecommand{\bibinfo}[2]{#2}
\providecommand{\eprint}[2][]{\url{#2}}

\bibitem[{\citenamefont{{Maekawa (Ed.)}}(2006)}]{Maekawa:2006}
\bibinfo{author}{\bibfnamefont{S.}~\bibnamefont{{Maekawa (Ed.)}}},
  \emph{\bibinfo{title}{Concepts in Spin Electronics}}
  (\bibinfo{publisher}{Oxford University Press}, \bibinfo{year}{2006}).

\bibitem[{\citenamefont{Maekawa and {T. Shinjo (Eds.)}}(2002)}]{Maekawa:2002B}
\bibinfo{author}{\bibfnamefont{S.}~\bibnamefont{Maekawa}} \bibnamefont{and}
  \bibinfo{author}{\bibnamefont{{T. Shinjo (Eds.)}}},
  \emph{\bibinfo{title}{Spin Dependent Transport in Magnetic Nanostructures}}
  (\bibinfo{publisher}{Taylor and Francis, New York}, \bibinfo{year}{2002}).

\bibitem[{\citenamefont{Parkin et~al.}(2003)\citenamefont{Parkin, Jiang,
  Kaiser, Panchula, Roche, and Samant}}]{Parkin2003:PIEEE}
\bibinfo{author}{\bibfnamefont{S.~S.~P.} \bibnamefont{Parkin}},
  \bibinfo{author}{\bibfnamefont{X.}~\bibnamefont{Jiang}},
  \bibinfo{author}{\bibfnamefont{C.}~\bibnamefont{Kaiser}},
  \bibinfo{author}{\bibfnamefont{A.}~\bibnamefont{Panchula}},
  \bibinfo{author}{\bibfnamefont{K.}~\bibnamefont{Roche}}, \bibnamefont{and}
  \bibinfo{author}{\bibfnamefont{M.}~\bibnamefont{Samant}},
  \bibinfo{journal}{Proc. IEEE} \textbf{\bibinfo{volume}{91}},
  \bibinfo{pages}{661} (\bibinfo{year}{2003}).

\bibitem[{\citenamefont{Prinz}(1998)}]{Prinz1998:S}
\bibinfo{author}{\bibfnamefont{G.}~\bibnamefont{Prinz}}, \bibinfo{journal}{{\sl
  Science}} \textbf{\bibinfo{volume}{282}}, \bibinfo{pages}{1660}
  (\bibinfo{year}{1998}).

\bibitem[{\citenamefont{Ansermet}(1998)}]{Ansermet1998:JPCM}
\bibinfo{author}{\bibfnamefont{J.-P.} \bibnamefont{Ansermet}},
  \bibinfo{journal}{J. Phys.: Condens. Matter} \textbf{\bibinfo{volume}{10}},
  \bibinfo{pages}{6027} (\bibinfo{year}{1998}).

\bibitem[{\citenamefont{{Hartman (Ed.)}}(2000)}]{Hartmann:2000}
\bibinfo{author}{\bibfnamefont{U.}~\bibnamefont{{Hartman (Ed.)}}},
  \emph{\bibinfo{title}{Magnetic Multilayers and Giant Magnetoresistance}}
  (\bibinfo{publisher}{Springer, Berlin}, \bibinfo{year}{2000}).

\bibitem[{\citenamefont{Hirota et~al.}(2002)\citenamefont{Hirota, Sakakima, and
  Inomata}}]{Hirota:2002}
\bibinfo{author}{\bibfnamefont{E.}~\bibnamefont{Hirota}},
  \bibinfo{author}{\bibfnamefont{H.}~\bibnamefont{Sakakima}}, \bibnamefont{and}
  \bibinfo{author}{\bibfnamefont{K.}~\bibnamefont{Inomata}},
  \emph{\bibinfo{title}{Giant Magneto-Resistance Devices}}
  (\bibinfo{publisher}{Springer, Berlin}, \bibinfo{year}{2002}).

\bibitem[{\citenamefont{Gregg et~al.}(1997)\citenamefont{Gregg, Allen, Viart,
  Kirschman, Sirisathitkul, Schille, Gester, Thompson, Sparks, {Da Costa}
  et~al.}}]{Gregg1997:JMMM}
\bibinfo{author}{\bibfnamefont{J.}~\bibnamefont{Gregg}},
  \bibinfo{author}{\bibfnamefont{W.}~\bibnamefont{Allen}},
  \bibinfo{author}{\bibfnamefont{N.}~\bibnamefont{Viart}},
  \bibinfo{author}{\bibfnamefont{R.}~\bibnamefont{Kirschman}},
  \bibinfo{author}{\bibfnamefont{C.}~\bibnamefont{Sirisathitkul}},
  \bibinfo{author}{\bibfnamefont{J.-P.} \bibnamefont{Schille}},
  \bibinfo{author}{\bibfnamefont{M.}~\bibnamefont{Gester}},
  \bibinfo{author}{\bibfnamefont{S.}~\bibnamefont{Thompson}},
  \bibinfo{author}{\bibfnamefont{P.}~\bibnamefont{Sparks}},
  \bibinfo{author}{\bibfnamefont{V.}~\bibnamefont{{Da Costa}}},
  \bibnamefont{et~al.}, \bibinfo{journal}{J. Magn. Magn. Mater.}
  \textbf{\bibinfo{volume}{175}}, \bibinfo{pages}{1} (\bibinfo{year}{1997}).

\bibitem[{\citenamefont{Johnson}(1994)}]{Johnson1994:IEEES}
\bibinfo{author}{\bibfnamefont{M.}~\bibnamefont{Johnson}},
  \bibinfo{journal}{IEEE Spectrum} \textbf{\bibinfo{volume}{31}},
  \bibinfo{pages}{47} (\bibinfo{year}{1994}).

\bibitem[{\citenamefont{Moodera and Mathon}(1999)}]{Moodera1999:JMMM}
\bibinfo{author}{\bibfnamefont{J.~S.} \bibnamefont{Moodera}} \bibnamefont{and}
  \bibinfo{author}{\bibfnamefont{G.}~\bibnamefont{Mathon}},
  \bibinfo{journal}{J. Mag. Magn. Mater.} \textbf{\bibinfo{volume}{200}},
  \bibinfo{pages}{248} (\bibinfo{year}{1999}).

\bibitem[{\citenamefont{Parkin et~al.}(1999)\citenamefont{Parkin, Roche,
  Samant, Rice, Beyers, Scheuerlein, {O'Sullivan}, Brown, Bucchigano, Abraham
  et~al.}}]{Parkin1999:JAP}
\bibinfo{author}{\bibfnamefont{S.~S.~P.} \bibnamefont{Parkin}},
  \bibinfo{author}{\bibfnamefont{K.~P.} \bibnamefont{Roche}},
  \bibinfo{author}{\bibfnamefont{M.~G.} \bibnamefont{Samant}},
  \bibinfo{author}{\bibfnamefont{P.~M.} \bibnamefont{Rice}},
  \bibinfo{author}{\bibfnamefont{R.~B.} \bibnamefont{Beyers}},
  \bibinfo{author}{\bibfnamefont{R.~E.} \bibnamefont{Scheuerlein}},
  \bibinfo{author}{\bibfnamefont{E.~J.} \bibnamefont{{O'Sullivan}}},
  \bibinfo{author}{\bibfnamefont{S.~L.} \bibnamefont{Brown}},
  \bibinfo{author}{\bibfnamefont{J.}~\bibnamefont{Bucchigano}},
  \bibinfo{author}{\bibfnamefont{D.~W.} \bibnamefont{Abraham}},
  \bibnamefont{et~al.}, \bibinfo{journal}{J. Appl. Phys.}
  \textbf{\bibinfo{volume}{85}}, \bibinfo{pages}{5828} (\bibinfo{year}{1999}).

\bibitem[{\citenamefont{Tehrani et~al.}(2000)\citenamefont{Tehrani, Engel,
  Slaughter, Chen, DeHerrera, Durlam, Naji, Whig, Janesky, and
  Calder}}]{Tehrani2000:IEEE}
\bibinfo{author}{\bibfnamefont{S.}~\bibnamefont{Tehrani}},
  \bibinfo{author}{\bibfnamefont{B.}~\bibnamefont{Engel}},
  \bibinfo{author}{\bibfnamefont{J.~M.} \bibnamefont{Slaughter}},
  \bibinfo{author}{\bibfnamefont{E.}~\bibnamefont{Chen}},
  \bibinfo{author}{\bibfnamefont{M.}~\bibnamefont{DeHerrera}},
  \bibinfo{author}{\bibfnamefont{M.}~\bibnamefont{Durlam}},
  \bibinfo{author}{\bibfnamefont{P.}~\bibnamefont{Naji}},
  \bibinfo{author}{\bibfnamefont{R.}~\bibnamefont{Whig}},
  \bibinfo{author}{\bibfnamefont{J.}~\bibnamefont{Janesky}}, \bibnamefont{and}
  \bibinfo{author}{\bibfnamefont{J.}~\bibnamefont{Calder}},
  \bibinfo{journal}{IEEE Trans. Magn.} \textbf{\bibinfo{volume}{36}},
  \bibinfo{pages}{2752} (\bibinfo{year}{2000}).

\bibitem[{\citenamefont{Johnson}(2001)}]{Johnson2001:JS}
\bibinfo{author}{\bibfnamefont{M.}~\bibnamefont{Johnson}}, \bibinfo{journal}{J.
  Supercond.} \textbf{\bibinfo{volume}{14}}, \bibinfo{pages}{273}
  (\bibinfo{year}{2001}).

\bibitem[{\citenamefont{{\v{Z}uti\'{c}}
  et~al.}(2004)\citenamefont{{\v{Z}uti\'{c}}, Fabian, and {Das
  Sarma}}}]{Zutic2004:RMP}
\bibinfo{author}{\bibfnamefont{I.}~\bibnamefont{{\v{Z}uti\'{c}}}},
  \bibinfo{author}{\bibfnamefont{J.}~\bibnamefont{Fabian}}, \bibnamefont{and}
  \bibinfo{author}{\bibfnamefont{S.}~\bibnamefont{{Das Sarma}}},
  \bibinfo{journal}{Rev. Mod. Phys.} \textbf{\bibinfo{volume}{76}},
  \bibinfo{pages}{323} (\bibinfo{year}{2004}).

\bibitem[{\citenamefont{Datta and Das}(1990)}]{Datta1990:APL}
\bibinfo{author}{\bibfnamefont{S.}~\bibnamefont{Datta}} \bibnamefont{and}
  \bibinfo{author}{\bibfnamefont{B.}~\bibnamefont{Das}},
  \bibinfo{journal}{Appl. Phys. Lett.} \textbf{\bibinfo{volume}{56}},
  \bibinfo{pages}{665} (\bibinfo{year}{1990}).

\bibitem[{\citenamefont{Bychkov and Rashba}(1984)}]{Bychkov1984:JPC}
\bibinfo{author}{\bibfnamefont{Y.~A.} \bibnamefont{Bychkov}} \bibnamefont{and}
  \bibinfo{author}{\bibfnamefont{E.~I.} \bibnamefont{Rashba}},
  \bibinfo{journal}{J. Phys. C} \textbf{\bibinfo{volume}{17}},
  \bibinfo{pages}{6039} (\bibinfo{year}{1984}).

\bibitem[{\citenamefont{Winkler}(2003)}]{Winkler:2003}
\bibinfo{author}{\bibfnamefont{R.}~\bibnamefont{Winkler}},
  \emph{\bibinfo{title}{Spin-Orbit Coupling Effecs in Two-Dimensional Electron
  and Hole Systems}} (\bibinfo{publisher}{Spinger, Berlin},
  \bibinfo{year}{2003}).

\bibitem[{\citenamefont{Wang et~al.}(2003)\citenamefont{Wang, Wang, and
  Guo}}]{Wang2002:P}
\bibinfo{author}{\bibfnamefont{B.}~\bibnamefont{Wang}},
  \bibinfo{author}{\bibfnamefont{J.}~\bibnamefont{Wang}}, \bibnamefont{and}
  \bibinfo{author}{\bibfnamefont{H.}~\bibnamefont{Guo}},
  \bibinfo{journal}{Phys. Rev. B} \textbf{\bibinfo{volume}{67}},
  \bibinfo{pages}{092408} (\bibinfo{year}{2003}).

\bibitem[{\citenamefont{Schliemann et~al.}(2003)\citenamefont{Schliemann,
  Egues, and Loss}}]{Schliemann2002:P}
\bibinfo{author}{\bibfnamefont{J.}~\bibnamefont{Schliemann}},
  \bibinfo{author}{\bibfnamefont{J.~C.} \bibnamefont{Egues}}, \bibnamefont{and}
  \bibinfo{author}{\bibfnamefont{D.}~\bibnamefont{Loss}},
  \bibinfo{journal}{Phys. Rev. Lett.} \textbf{\bibinfo{volume}{90}},
  \bibinfo{pages}{146801} (\bibinfo{year}{2003}).

\bibitem[{\citenamefont{Matsuyama et~al.}(2002)\citenamefont{Matsuyama, Hu,
  Grundler, Meier, and Merkt}}]{Matsuyama2002:PRB}
\bibinfo{author}{\bibfnamefont{T.}~\bibnamefont{Matsuyama}},
  \bibinfo{author}{\bibfnamefont{C.-M.} \bibnamefont{Hu}},
  \bibinfo{author}{\bibfnamefont{D.}~\bibnamefont{Grundler}},
  \bibinfo{author}{\bibfnamefont{G.}~\bibnamefont{Meier}}, \bibnamefont{and}
  \bibinfo{author}{\bibfnamefont{U.}~\bibnamefont{Merkt}},
  \bibinfo{journal}{Phys. Rev. B} \textbf{\bibinfo{volume}{65}},
  \bibinfo{pages}{155322} (\bibinfo{year}{2002}).

\bibitem[{\citenamefont{Mirales and Kirczenow}(2001)}]{Mirales2001:PRB}
\bibinfo{author}{\bibfnamefont{F.}~\bibnamefont{Mirales}} \bibnamefont{and}
  \bibinfo{author}{\bibfnamefont{G.}~\bibnamefont{Kirczenow}},
  \bibinfo{journal}{Phys. Rev. B} \textbf{\bibinfo{volume}{64}},
  \bibinfo{pages}{024426} (\bibinfo{year}{2001}).

\bibitem[{\citenamefont{Ciuti et~al.}(2002)\citenamefont{Ciuti, McGuire, and
  Sham}}]{Ciuti2002:APL}
\bibinfo{author}{\bibfnamefont{C.}~\bibnamefont{Ciuti}},
  \bibinfo{author}{\bibfnamefont{J.~P.} \bibnamefont{McGuire}},
  \bibnamefont{and} \bibinfo{author}{\bibfnamefont{L.~J.} \bibnamefont{Sham}},
  \bibinfo{journal}{Appl. Phys. Lett.} \textbf{\bibinfo{volume}{81}},
  \bibinfo{pages}{4781} (\bibinfo{year}{2002}).

\bibitem[{\citenamefont{Nikonov and Bourianoff}(2005)}]{Nikonov2005:IEEETN}
\bibinfo{author}{\bibfnamefont{D.~E.} \bibnamefont{Nikonov}} \bibnamefont{and}
  \bibinfo{author}{\bibfnamefont{G.~I.} \bibnamefont{Bourianoff}},
  \bibinfo{journal}{IEEE Trans. Nanotech.} \textbf{\bibinfo{volume}{4}},
  \bibinfo{pages}{206} (\bibinfo{year}{2005}).

\bibitem[{\citenamefont{Sugahara and Tanaka}(2004)}]{Sugahara2004:APL}
\bibinfo{author}{\bibfnamefont{S.}~\bibnamefont{Sugahara}} \bibnamefont{and}
  \bibinfo{author}{\bibfnamefont{M.}~\bibnamefont{Tanaka}},
  \bibinfo{journal}{Appl. Phys. Lett.} \textbf{\bibinfo{volume}{84}},
  \bibinfo{pages}{2307} (\bibinfo{year}{2004}).

\bibitem[{\citenamefont{Sugahara and Tanaka}(2005)}]{Sugahara2005:JAP}
\bibinfo{author}{\bibfnamefont{S.}~\bibnamefont{Sugahara}} \bibnamefont{and}
  \bibinfo{author}{\bibfnamefont{M.}~\bibnamefont{Tanaka}},
  \bibinfo{journal}{J. Appl. Phys.} \textbf{\bibinfo{volume}{97}},
  \bibinfo{pages}{10D503} (\bibinfo{year}{2005}).

\bibitem[{\citenamefont{Sch{\"a}pers et~al.}(2001)\citenamefont{Sch{\"a}pers,
  Nitta, h.~B.~Heersche, and Takayanagi}}]{Schapers2001:PRB}
\bibinfo{author}{\bibfnamefont{T.}~\bibnamefont{Sch{\"a}pers}},
  \bibinfo{author}{\bibfnamefont{J.}~\bibnamefont{Nitta}},
  \bibinfo{author}{\bibnamefont{h.~B.~Heersche}}, \bibnamefont{and}
  \bibinfo{author}{\bibfnamefont{T.}~\bibnamefont{Takayanagi}},
  \bibinfo{journal}{Phys. Rev. B} \textbf{\bibinfo{volume}{64}},
  \bibinfo{pages}{125314} (\bibinfo{year}{2001}).

\bibitem[{\citenamefont{Sahoo et~al.}(2005)\citenamefont{Sahoo, Kontos, annd
  C.~Hoffman, Gr{\"a}ber, Cottet, and Sch{\"o}nenberger}}]{Sahoo2005:NP}
\bibinfo{author}{\bibfnamefont{S.}~\bibnamefont{Sahoo}},
  \bibinfo{author}{\bibfnamefont{T.}~\bibnamefont{Kontos}},
  \bibinfo{author}{\bibfnamefont{J.~F.} \bibnamefont{annd C.~Hoffman}},
  \bibinfo{author}{\bibfnamefont{M.}~\bibnamefont{Gr{\"a}ber}},
  \bibinfo{author}{\bibfnamefont{A.}~\bibnamefont{Cottet}}, \bibnamefont{and}
  \bibinfo{author}{\bibfnamefont{C.}~\bibnamefont{Sch{\"o}nenberger}},
  \bibinfo{journal}{Nature Phys.} \textbf{\bibinfo{volume}{1}},
  \bibinfo{pages}{99} (\bibinfo{year}{2005}).

\bibitem[{\citenamefont{\v{Z}uti\'{c} and Fuhrer}(2005)}]{Zutic2005:NP}
\bibinfo{author}{\bibfnamefont{I.}~\bibnamefont{\v{Z}uti\'{c}}}
  \bibnamefont{and} \bibinfo{author}{\bibfnamefont{M.}~\bibnamefont{Fuhrer}},
  \bibinfo{journal}{Nature Phys.} \textbf{\bibinfo{volume}{1}},
  \bibinfo{pages}{85} (\bibinfo{year}{2005}).

\bibitem[{bip()}]{bipolar}
\bibinfo{note}{This is a conventional meaning of the term bipolar, as used in
  the physics of semiconductors. However, the term {\it bipolar} has also been
  used to describe an analogy between the coexistence of two spin carrier
  populations (of spin up and spin down) in spin-polarized transport and two
  charge carrier populations (electrons and holes) in bipolar charge transport.
  See, for example, M. Johnson, Science {\bf 260}, 320 (1993).}

\bibitem[{\citenamefont{Maekawa et~al.}(2002)\citenamefont{Maekawa, Takahashi,
  and Imamura}}]{Maekawa:2002}
\bibinfo{author}{\bibfnamefont{S.}~\bibnamefont{Maekawa}},
  \bibinfo{author}{\bibfnamefont{S.}~\bibnamefont{Takahashi}},
  \bibnamefont{and} \bibinfo{author}{\bibfnamefont{H.}~\bibnamefont{Imamura}},
  in \emph{\bibinfo{booktitle}{Spin Dependent Transport in Magnetic
  Nanostructures}}, edited by
  \bibinfo{editor}{\bibfnamefont{S.}~\bibnamefont{Maekawa}} \bibnamefont{and}
  \bibinfo{editor}{\bibfnamefont{T.}~\bibnamefont{Shinjo}}
  (\bibinfo{publisher}{Taylor and Francis, New York}, \bibinfo{year}{2002}),
  pp. \bibinfo{pages}{143--236}.

\bibitem[{\citenamefont{Shockley}(1950)}]{Shockley:1950}
\bibinfo{author}{\bibfnamefont{W.}~\bibnamefont{Shockley}},
  \emph{\bibinfo{title}{Electrons and Holes in Semiconductors}}
  (\bibinfo{publisher}{D. {Van Nostrand}, Princeton}, \bibinfo{year}{1950}).

\bibitem[{\citenamefont{Sze}(1981)}]{Sze:1981}
\bibinfo{author}{\bibfnamefont{S.~M.} \bibnamefont{Sze}},
  \emph{\bibinfo{title}{Physics of {Semiconductor} {Devices}}}
  (\bibinfo{publisher}{John Wiley, New York}, \bibinfo{year}{1981}).

\bibitem[{\citenamefont{Fiederling et~al.}(1999)\citenamefont{Fiederling,
  Kleim, Reuscher, Ossau, Schmidt, Waag, and Molenkamp}}]{Fiederling1999:N}
\bibinfo{author}{\bibfnamefont{R.}~\bibnamefont{Fiederling}},
  \bibinfo{author}{\bibfnamefont{M.}~\bibnamefont{Kleim}},
  \bibinfo{author}{\bibfnamefont{G.}~\bibnamefont{Reuscher}},
  \bibinfo{author}{\bibfnamefont{W.}~\bibnamefont{Ossau}},
  \bibinfo{author}{\bibfnamefont{G.}~\bibnamefont{Schmidt}},
  \bibinfo{author}{\bibfnamefont{A.}~\bibnamefont{Waag}}, \bibnamefont{and}
  \bibinfo{author}{\bibfnamefont{L.~W.} \bibnamefont{Molenkamp}},
  \bibinfo{journal}{{\sl Nature}} \textbf{\bibinfo{volume}{402}},
  \bibinfo{pages}{787} (\bibinfo{year}{1999}).

\bibitem[{\citenamefont{Jonker et~al.}(2000)\citenamefont{Jonker, Park,
  Bennett, Cheong, Kioseoglou, and Petrou}}]{Jonker2000:PRB}
\bibinfo{author}{\bibfnamefont{B.~T.} \bibnamefont{Jonker}},
  \bibinfo{author}{\bibfnamefont{Y.~D.} \bibnamefont{Park}},
  \bibinfo{author}{\bibfnamefont{B.~R.} \bibnamefont{Bennett}},
  \bibinfo{author}{\bibfnamefont{H.~D.} \bibnamefont{Cheong}},
  \bibinfo{author}{\bibfnamefont{G.}~\bibnamefont{Kioseoglou}},
  \bibnamefont{and} \bibinfo{author}{\bibfnamefont{A.}~\bibnamefont{Petrou}},
  \bibinfo{journal}{Phys. Rev. B} \textbf{\bibinfo{volume}{62}},
  \bibinfo{pages}{8180} (\bibinfo{year}{2000}).

\bibitem[{\citenamefont{Young et~al.}(2002)\citenamefont{Young,
  {Johnston-Halperin}, Awschalom, Ohno, and Ohno}}]{Young2002:APL}
\bibinfo{author}{\bibfnamefont{D.~K.} \bibnamefont{Young}},
  \bibinfo{author}{\bibfnamefont{E.}~\bibnamefont{{Johnston-Halperin}}},
  \bibinfo{author}{\bibfnamefont{D.~D.} \bibnamefont{Awschalom}},
  \bibinfo{author}{\bibfnamefont{Y.}~\bibnamefont{Ohno}}, \bibnamefont{and}
  \bibinfo{author}{\bibfnamefont{H.}~\bibnamefont{Ohno}},
  \bibinfo{journal}{Appl. Phys. Lett.} \textbf{\bibinfo{volume}{80}},
  \bibinfo{pages}{1598} (\bibinfo{year}{2002}).

\bibitem[{\citenamefont{Hanbicki et~al.}(2003)\citenamefont{Hanbicki, {van
  t'Erve}, Magno, Kioseoglou, Li, Jonker, Itskos, Mallory, Yasar, and
  Petrou}}]{Hanbicki2003:P}
\bibinfo{author}{\bibfnamefont{A.}~\bibnamefont{Hanbicki}},
  \bibinfo{author}{\bibfnamefont{O.~M.~J.} \bibnamefont{{van t'Erve}}},
  \bibinfo{author}{\bibfnamefont{R.}~\bibnamefont{Magno}},
  \bibinfo{author}{\bibfnamefont{G.}~\bibnamefont{Kioseoglou}},
  \bibinfo{author}{\bibfnamefont{C.~H.} \bibnamefont{Li}},
  \bibinfo{author}{\bibfnamefont{B.~T.} \bibnamefont{Jonker}},
  \bibinfo{author}{\bibfnamefont{G.}~\bibnamefont{Itskos}},
  \bibinfo{author}{\bibfnamefont{R.}~\bibnamefont{Mallory}},
  \bibinfo{author}{\bibfnamefont{M.}~\bibnamefont{Yasar}}, \bibnamefont{and}
  \bibinfo{author}{\bibfnamefont{A.}~\bibnamefont{Petrou}},
  \bibinfo{journal}{Appl. Phys. Lett.} \textbf{\bibinfo{volume}{82}},
  \bibinfo{pages}{4092} (\bibinfo{year}{2003}).

\bibitem[{\citenamefont{Jiang et~al.}(2003)\citenamefont{Jiang, Wang, van
  Dijken, Shelby, Macfarlane, Solomon, Harris, and Parkin}}]{Jiang2003:PRL}
\bibinfo{author}{\bibfnamefont{X.}~\bibnamefont{Jiang}},
  \bibinfo{author}{\bibfnamefont{R.}~\bibnamefont{Wang}},
  \bibinfo{author}{\bibfnamefont{S.}~\bibnamefont{van Dijken}},
  \bibinfo{author}{\bibfnamefont{R.}~\bibnamefont{Shelby}},
  \bibinfo{author}{\bibfnamefont{R.}~\bibnamefont{Macfarlane}},
  \bibinfo{author}{\bibfnamefont{G.~S.} \bibnamefont{Solomon}},
  \bibinfo{author}{\bibfnamefont{J.}~\bibnamefont{Harris}}, \bibnamefont{and}
  \bibinfo{author}{\bibfnamefont{S.~S.~P.} \bibnamefont{Parkin}},
  \bibinfo{journal}{Phys. Rev. Lett.} \textbf{\bibinfo{volume}{90}},
  \bibinfo{pages}{256603} (\bibinfo{year}{2003}).

\bibitem[{\citenamefont{{Van Roy} et~al.}(2006)\citenamefont{{Van Roy}, {Van
  Dorpe}, {De Boeck}, and Borghs}}]{VanRoy2006:MSEB}
\bibinfo{author}{\bibfnamefont{W.}~\bibnamefont{{Van Roy}}},
  \bibinfo{author}{\bibfnamefont{P.}~\bibnamefont{{Van Dorpe}}},
  \bibinfo{author}{\bibfnamefont{J.}~\bibnamefont{{De Boeck}}},
  \bibnamefont{and} \bibinfo{author}{\bibfnamefont{G.}~\bibnamefont{Borghs}},
  \bibinfo{journal}{Mater. Sci. Eng. B} \textbf{\bibinfo{volume}{126}},
  \bibinfo{pages}{155} (\bibinfo{year}{2006}).

\bibitem[{\citenamefont{Janik and Karczewski}(1988)}]{Janik1988:APPA}
\bibinfo{author}{\bibfnamefont{E.}~\bibnamefont{Janik}} \bibnamefont{and}
  \bibinfo{author}{\bibfnamefont{G.}~\bibnamefont{Karczewski}},
  \bibinfo{journal}{Acta Physica Polonica A} \textbf{\bibinfo{volume}{73}},
  \bibinfo{pages}{439} (\bibinfo{year}{1988}).

\bibitem[{\citenamefont{Munekata et~al.}(1989)\citenamefont{Munekata, Ohno, von
  Moln\'{a}r, Segm{\"u}ller, Chang, and Esaki}}]{Munekata1989:PRL}
\bibinfo{author}{\bibfnamefont{H.}~\bibnamefont{Munekata}},
  \bibinfo{author}{\bibfnamefont{H.}~\bibnamefont{Ohno}},
  \bibinfo{author}{\bibfnamefont{S.}~\bibnamefont{von Moln\'{a}r}},
  \bibinfo{author}{\bibfnamefont{A.}~\bibnamefont{Segm{\"u}ller}},
  \bibinfo{author}{\bibfnamefont{L.~L.} \bibnamefont{Chang}}, \bibnamefont{and}
  \bibinfo{author}{\bibfnamefont{L.}~\bibnamefont{Esaki}},
  \bibinfo{journal}{Phys. Rev. Lett.} \textbf{\bibinfo{volume}{63}},
  \bibinfo{pages}{1849} (\bibinfo{year}{1989}).

\bibitem[{\citenamefont{Ohno et~al.}(1992)\citenamefont{Ohno, Munekata, Penney,
  von Moln\'{a}r, and Chang}}]{Ohno1992:PRL}
\bibinfo{author}{\bibfnamefont{H.}~\bibnamefont{Ohno}},
  \bibinfo{author}{\bibfnamefont{H.}~\bibnamefont{Munekata}},
  \bibinfo{author}{\bibfnamefont{T.}~\bibnamefont{Penney}},
  \bibinfo{author}{\bibfnamefont{S.}~\bibnamefont{von Moln\'{a}r}},
  \bibnamefont{and} \bibinfo{author}{\bibfnamefont{L.~L.} \bibnamefont{Chang}},
  \bibinfo{journal}{Phys. Rev. Lett.} \textbf{\bibinfo{volume}{68}},
  \bibinfo{pages}{2664} (\bibinfo{year}{1992}).

\bibitem[{\citenamefont{Munekata et~al.}(1991)\citenamefont{Munekata, Ohno,
  Ruf, Gambino, and Chang}}]{Munekata1991:JCG}
\bibinfo{author}{\bibfnamefont{H.}~\bibnamefont{Munekata}},
  \bibinfo{author}{\bibfnamefont{H.}~\bibnamefont{Ohno}},
  \bibinfo{author}{\bibfnamefont{R.~R.} \bibnamefont{Ruf}},
  \bibinfo{author}{\bibfnamefont{R.~J.} \bibnamefont{Gambino}},
  \bibnamefont{and} \bibinfo{author}{\bibfnamefont{L.~L.} \bibnamefont{Chang}},
  \bibinfo{journal}{J. Cryst. Growth} \textbf{\bibinfo{volume}{111}},
  \bibinfo{pages}{1011} (\bibinfo{year}{1991}).

\bibitem[{\citenamefont{Ohno et~al.}(1996)\citenamefont{Ohno, Shen, Matsukura,
  Oiwa, End, Katsumoto, and Iye}}]{Ohno1996:APL}
\bibinfo{author}{\bibfnamefont{H.}~\bibnamefont{Ohno}},
  \bibinfo{author}{\bibfnamefont{A.}~\bibnamefont{Shen}},
  \bibinfo{author}{\bibfnamefont{F.}~\bibnamefont{Matsukura}},
  \bibinfo{author}{\bibfnamefont{A.}~\bibnamefont{Oiwa}},
  \bibinfo{author}{\bibfnamefont{A.}~\bibnamefont{End}},
  \bibinfo{author}{\bibfnamefont{S.}~\bibnamefont{Katsumoto}},
  \bibnamefont{and} \bibinfo{author}{\bibfnamefont{Y.}~\bibnamefont{Iye}},
  \bibinfo{journal}{Appl. Phys. Lett.} \textbf{\bibinfo{volume}{69}},
  \bibinfo{pages}{363} (\bibinfo{year}{1996}).

\bibitem[{\citenamefont{{Van Esch} et~al.}(1997)\citenamefont{{Van Esch}, {Van
  Bockstal}, {De Boeck}, Verbanck, {van Steenbergen}, Wellmann, Grietens,
  Bogaerts, Herlach, and Borghs}}]{VanEsch1997:PRB}
\bibinfo{author}{\bibfnamefont{A.}~\bibnamefont{{Van Esch}}},
  \bibinfo{author}{\bibfnamefont{L.}~\bibnamefont{{Van Bockstal}}},
  \bibinfo{author}{\bibfnamefont{J.}~\bibnamefont{{De Boeck}}},
  \bibinfo{author}{\bibfnamefont{G.}~\bibnamefont{Verbanck}},
  \bibinfo{author}{\bibfnamefont{A.~S.} \bibnamefont{{van Steenbergen}}},
  \bibinfo{author}{\bibfnamefont{P.~J.} \bibnamefont{Wellmann}},
  \bibinfo{author}{\bibfnamefont{B.}~\bibnamefont{Grietens}},
  \bibinfo{author}{\bibfnamefont{R.}~\bibnamefont{Bogaerts}},
  \bibinfo{author}{\bibfnamefont{F.}~\bibnamefont{Herlach}}, \bibnamefont{and}
  \bibinfo{author}{\bibfnamefont{G.}~\bibnamefont{Borghs}},
  \bibinfo{journal}{Phys. Rev. B} \textbf{\bibinfo{volume}{56}},
  \bibinfo{pages}{13103} (\bibinfo{year}{1997}).

\bibitem[{\citenamefont{Hayashi et~al.}(1997)\citenamefont{Hayashi, Tanaka,
  Nishinaga, Shimada, Tsuchiya, and Otuka}}]{Hayashi1997:JCG}
\bibinfo{author}{\bibfnamefont{T.}~\bibnamefont{Hayashi}},
  \bibinfo{author}{\bibfnamefont{M.}~\bibnamefont{Tanaka}},
  \bibinfo{author}{\bibfnamefont{T.}~\bibnamefont{Nishinaga}},
  \bibinfo{author}{\bibfnamefont{H.}~\bibnamefont{Shimada}},
  \bibinfo{author}{\bibfnamefont{T.}~\bibnamefont{Tsuchiya}}, \bibnamefont{and}
  \bibinfo{author}{\bibfnamefont{Y.}~\bibnamefont{Otuka}}, \bibinfo{journal}{J.
  Cryst. Growth} \textbf{\bibinfo{volume}{175/176}}, \bibinfo{pages}{1063}
  (\bibinfo{year}{1997}).

\bibitem[{\citenamefont{Ohno}(1998)}]{Ohno1998:S}
\bibinfo{author}{\bibfnamefont{H.}~\bibnamefont{Ohno}}, \bibinfo{journal}{{\sl
  Science}} \textbf{\bibinfo{volume}{281}}, \bibinfo{pages}{951}
  (\bibinfo{year}{1998}).

\bibitem[{\citenamefont{Dietl}(2002)}]{Dietl2002:SST}
\bibinfo{author}{\bibfnamefont{T.}~\bibnamefont{Dietl}},
  \bibinfo{journal}{Semicond. Sci. Technol.} \textbf{\bibinfo{volume}{17}},
  \bibinfo{pages}{377} (\bibinfo{year}{2002}).

\bibitem[{\citenamefont{Jungwirth et~al.}(2006)\citenamefont{Jungwirth, Sinova,
  Masek, Kucera, and MacDonald}}]{Jungwirth2006:P}
\bibinfo{author}{\bibfnamefont{T.}~\bibnamefont{Jungwirth}},
  \bibinfo{author}{\bibfnamefont{J.}~\bibnamefont{Sinova}},
  \bibinfo{author}{\bibfnamefont{J.}~\bibnamefont{Masek}},
  \bibinfo{author}{\bibfnamefont{J.}~\bibnamefont{Kucera}}, \bibnamefont{and}
  \bibinfo{author}{\bibfnamefont{A.~H.} \bibnamefont{MacDonald}}
  (\bibinfo{year}{2006}), \eprint{cond-mat/0603080}.

\bibitem[{\citenamefont{Ohno et~al.}(2000{\natexlab{a}})\citenamefont{Ohno,
  Arata, Matsukura, Ohtani, Wang, and Ohno}}]{Ohno2000:ASS}
\bibinfo{author}{\bibfnamefont{Y.}~\bibnamefont{Ohno}},
  \bibinfo{author}{\bibfnamefont{I.}~\bibnamefont{Arata}},
  \bibinfo{author}{\bibfnamefont{F.}~\bibnamefont{Matsukura}},
  \bibinfo{author}{\bibfnamefont{K.}~\bibnamefont{Ohtani}},
  \bibinfo{author}{\bibfnamefont{S.}~\bibnamefont{Wang}}, \bibnamefont{and}
  \bibinfo{author}{\bibfnamefont{H.}~\bibnamefont{Ohno}},
  \bibinfo{journal}{Appl. Surf. Sci.} \textbf{\bibinfo{volume}{159-160}},
  \bibinfo{pages}{308} (\bibinfo{year}{2000}{\natexlab{a}}).

\bibitem[{\citenamefont{Kohda et~al.}(2001)\citenamefont{Kohda, Ohno, Takamura,
  Matsukura, and Ohno}}]{Kohda2001:JJAP}
\bibinfo{author}{\bibfnamefont{M.}~\bibnamefont{Kohda}},
  \bibinfo{author}{\bibfnamefont{Y.}~\bibnamefont{Ohno}},
  \bibinfo{author}{\bibfnamefont{K.}~\bibnamefont{Takamura}},
  \bibinfo{author}{\bibfnamefont{F.}~\bibnamefont{Matsukura}},
  \bibnamefont{and} \bibinfo{author}{\bibfnamefont{H.}~\bibnamefont{Ohno}},
  \bibinfo{journal}{Jpn. J. Appl. Phys.} \textbf{\bibinfo{volume}{40}},
  \bibinfo{pages}{L1274} (\bibinfo{year}{2001}).

\bibitem[{\citenamefont{Johnston-Halperin
  et~al.}(2002)\citenamefont{Johnston-Halperin, Lofgreen, Kawakami, Young,
  Coldren, Gossard, and Awschalom}}]{Johnston-Halperin2002:PRB}
\bibinfo{author}{\bibfnamefont{E.}~\bibnamefont{Johnston-Halperin}},
  \bibinfo{author}{\bibfnamefont{D.}~\bibnamefont{Lofgreen}},
  \bibinfo{author}{\bibfnamefont{R.~K.} \bibnamefont{Kawakami}},
  \bibinfo{author}{\bibfnamefont{D.~K.} \bibnamefont{Young}},
  \bibinfo{author}{\bibfnamefont{L.}~\bibnamefont{Coldren}},
  \bibinfo{author}{\bibfnamefont{A.~C.} \bibnamefont{Gossard}},
  \bibnamefont{and} \bibinfo{author}{\bibfnamefont{D.~D.}
  \bibnamefont{Awschalom}}, \bibinfo{journal}{Phys. Rev. B}
  \textbf{\bibinfo{volume}{65}}, \bibinfo{pages}{041306}
  (\bibinfo{year}{2002}).

\bibitem[{\citenamefont{Arata et~al.}(2001)\citenamefont{Arata, Ohno,
  Matsukura, and Ohno}}]{Arata2001:PE}
\bibinfo{author}{\bibfnamefont{I.}~\bibnamefont{Arata}},
  \bibinfo{author}{\bibfnamefont{Y.}~\bibnamefont{Ohno}},
  \bibinfo{author}{\bibfnamefont{F.}~\bibnamefont{Matsukura}},
  \bibnamefont{and} \bibinfo{author}{\bibfnamefont{H.}~\bibnamefont{Ohno}},
  \bibinfo{journal}{Physica E} \textbf{\bibinfo{volume}{10}},
  \bibinfo{pages}{288} (\bibinfo{year}{2001}).

\bibitem[{\citenamefont{{Van Dorpe} et~al.}(2004)\citenamefont{{Van Dorpe},
  Liu, Roy, Motsnyi, Sawicki, Borghs, and {De Boeck}}}]{vanDorpe2003:P}
\bibinfo{author}{\bibfnamefont{P.}~\bibnamefont{{Van Dorpe}}},
  \bibinfo{author}{\bibfnamefont{Z.}~\bibnamefont{Liu}},
  \bibinfo{author}{\bibfnamefont{W.~V.} \bibnamefont{Roy}},
  \bibinfo{author}{\bibfnamefont{V.~F.} \bibnamefont{Motsnyi}},
  \bibinfo{author}{\bibfnamefont{M.}~\bibnamefont{Sawicki}},
  \bibinfo{author}{\bibfnamefont{G.}~\bibnamefont{Borghs}}, \bibnamefont{and}
  \bibinfo{author}{\bibfnamefont{J.}~\bibnamefont{{De Boeck}}},
  \bibinfo{journal}{Appl. Phys. Lett.} \textbf{\bibinfo{volume}{84}},
  \bibinfo{pages}{3495} (\bibinfo{year}{2004}).

\bibitem[{\citenamefont{Tsui et~al.}(2003)\citenamefont{Tsui, Ma, and
  He}}]{Tsui2003:APL}
\bibinfo{author}{\bibfnamefont{F.}~\bibnamefont{Tsui}},
  \bibinfo{author}{\bibfnamefont{L.}~\bibnamefont{Ma}}, \bibnamefont{and}
  \bibinfo{author}{\bibfnamefont{L.}~\bibnamefont{He}}, \bibinfo{journal}{Appl.
  Phys. Lett.} \textbf{\bibinfo{volume}{83}}, \bibinfo{pages}{954}
  (\bibinfo{year}{2003}).

\bibitem[{\citenamefont{Dietl}(2003)}]{Dietl2003:NM}
\bibinfo{author}{\bibfnamefont{T.}~\bibnamefont{Dietl}},
  \bibinfo{journal}{Nature Mater.} \textbf{\bibinfo{volume}{2}},
  \bibinfo{pages}{646} (\bibinfo{year}{2003}).

\bibitem[{\citenamefont{Samarth et~al.}(2003)\citenamefont{Samarth, Chun, Ku,
  Potashnik, and Schiffer}}]{Samarth2003:SSC}
\bibinfo{author}{\bibfnamefont{N.}~\bibnamefont{Samarth}},
  \bibinfo{author}{\bibfnamefont{S.~H.} \bibnamefont{Chun}},
  \bibinfo{author}{\bibfnamefont{K.~C.} \bibnamefont{Ku}},
  \bibinfo{author}{\bibfnamefont{S.~J.} \bibnamefont{Potashnik}},
  \bibnamefont{and} \bibinfo{author}{\bibfnamefont{P.}~\bibnamefont{Schiffer}},
  \bibinfo{journal}{Solid State Commun.} \textbf{\bibinfo{volume}{127}},
  \bibinfo{pages}{173} (\bibinfo{year}{2003}).

\bibitem[{\citenamefont{Park et~al.}(2002)\citenamefont{Park, Hanbicki, Erwin,
  Hellberg, Sullivan, Mattson, Ambrose, Wilson, Spanos, and
  Jonker}}]{Park2002:S}
\bibinfo{author}{\bibfnamefont{Y.~D.} \bibnamefont{Park}},
  \bibinfo{author}{\bibfnamefont{A.~T.} \bibnamefont{Hanbicki}},
  \bibinfo{author}{\bibfnamefont{S.~C.} \bibnamefont{Erwin}},
  \bibinfo{author}{\bibfnamefont{C.~S.} \bibnamefont{Hellberg}},
  \bibinfo{author}{\bibfnamefont{J.~M.} \bibnamefont{Sullivan}},
  \bibinfo{author}{\bibfnamefont{J.~E.} \bibnamefont{Mattson}},
  \bibinfo{author}{\bibfnamefont{T.~F.} \bibnamefont{Ambrose}},
  \bibinfo{author}{\bibfnamefont{A.}~\bibnamefont{Wilson}},
  \bibinfo{author}{\bibfnamefont{G.}~\bibnamefont{Spanos}}, \bibnamefont{and}
  \bibinfo{author}{\bibfnamefont{B.~T.} \bibnamefont{Jonker}},
  \bibinfo{journal}{{\sl Science}} \textbf{\bibinfo{volume}{295}},
  \bibinfo{pages}{651} (\bibinfo{year}{2002}).

\bibitem[{\citenamefont{Li et~al.}(2005)\citenamefont{Li, Shen, Thompson, and
  Weitering}}]{Li2005:APL}
\bibinfo{author}{\bibfnamefont{A.~P.} \bibnamefont{Li}},
  \bibinfo{author}{\bibfnamefont{J.}~\bibnamefont{Shen}},
  \bibinfo{author}{\bibfnamefont{J.~R.} \bibnamefont{Thompson}},
  \bibnamefont{and} \bibinfo{author}{\bibfnamefont{H.~H.}
  \bibnamefont{Weitering}}, \bibinfo{journal}{Appl. Phys. Lett.}
  \textbf{\bibinfo{volume}{86}}, \bibinfo{pages}{152507}
  (\bibinfo{year}{2005}).

\bibitem[{\citenamefont{Koshihara et~al.}(1997)\citenamefont{Koshihara, Oiwa,
  Hirasawa, Katsumoto, Iye, Urano, Takagi, and Munekata}}]{Koshihara1997:PRL}
\bibinfo{author}{\bibfnamefont{S.}~\bibnamefont{Koshihara}},
  \bibinfo{author}{\bibfnamefont{A.}~\bibnamefont{Oiwa}},
  \bibinfo{author}{\bibfnamefont{M.}~\bibnamefont{Hirasawa}},
  \bibinfo{author}{\bibfnamefont{S.}~\bibnamefont{Katsumoto}},
  \bibinfo{author}{\bibfnamefont{Y.}~\bibnamefont{Iye}},
  \bibinfo{author}{\bibfnamefont{S.}~\bibnamefont{Urano}},
  \bibinfo{author}{\bibfnamefont{H.}~\bibnamefont{Takagi}}, \bibnamefont{and}
  \bibinfo{author}{\bibfnamefont{H.}~\bibnamefont{Munekata}},
  \bibinfo{journal}{Phys. Rev. Lett.} \textbf{\bibinfo{volume}{78}},
  \bibinfo{pages}{4617} (\bibinfo{year}{1997}).

\bibitem[{\citenamefont{Oiwa et~al.}(2002)\citenamefont{Oiwa, Mitsumori,
  Moriya, Supinski, and Munekata}}]{Oiwa2002:PRL}
\bibinfo{author}{\bibfnamefont{A.}~\bibnamefont{Oiwa}},
  \bibinfo{author}{\bibfnamefont{Y.}~\bibnamefont{Mitsumori}},
  \bibinfo{author}{\bibfnamefont{R.}~\bibnamefont{Moriya}},
  \bibinfo{author}{\bibfnamefont{T.}~\bibnamefont{Supinski}}, \bibnamefont{and}
  \bibinfo{author}{\bibfnamefont{H.}~\bibnamefont{Munekata}},
  \bibinfo{journal}{Phys. Rev. Lett.} \textbf{\bibinfo{volume}{88}},
  \bibinfo{pages}{137202} (\bibinfo{year}{2002}).

\bibitem[{\citenamefont{Ohno et~al.}(2000{\natexlab{b}})\citenamefont{Ohno,
  Chiba, Matsukura, Abe, Dietl, Ohno, and Ohtani}}]{Ohno2000:N}
\bibinfo{author}{\bibfnamefont{H.}~\bibnamefont{Ohno}},
  \bibinfo{author}{\bibfnamefont{D.}~\bibnamefont{Chiba}},
  \bibinfo{author}{\bibfnamefont{F.}~\bibnamefont{Matsukura}},
  \bibinfo{author}{\bibfnamefont{T.~O.~E.} \bibnamefont{Abe}},
  \bibinfo{author}{\bibfnamefont{T.}~\bibnamefont{Dietl}},
  \bibinfo{author}{\bibfnamefont{Y.}~\bibnamefont{Ohno}}, \bibnamefont{and}
  \bibinfo{author}{\bibfnamefont{K.}~\bibnamefont{Ohtani}},
  \bibinfo{journal}{{\sl Nature}} \textbf{\bibinfo{volume}{408}},
  \bibinfo{pages}{944} (\bibinfo{year}{2000}{\natexlab{b}}).

\bibitem[{\citenamefont{{De Boeck} et~al.}(1996)\citenamefont{{De Boeck},
  Oesterholt, {Van Esch}, r, Bruynseraede, {Van Hoof}, and
  Borghs}}]{DeBoeck1996:APL}
\bibinfo{author}{\bibfnamefont{J.}~\bibnamefont{{De Boeck}}},
  \bibinfo{author}{\bibfnamefont{R.}~\bibnamefont{Oesterholt}},
  \bibinfo{author}{\bibfnamefont{A.}~\bibnamefont{{Van Esch}}},
  \bibinfo{author}{\bibfnamefont{H.~B.} \bibnamefont{r}},
  \bibinfo{author}{\bibfnamefont{C.}~\bibnamefont{Bruynseraede}},
  \bibinfo{author}{\bibfnamefont{C.}~\bibnamefont{{Van Hoof}}},
  \bibnamefont{and} \bibinfo{author}{\bibfnamefont{G.}~\bibnamefont{Borghs}},
  \bibinfo{journal}{App. Phys. Lett.} \textbf{\bibinfo{volume}{68}},
  \bibinfo{pages}{2744} (\bibinfo{year}{1996}).

\bibitem[{\citenamefont{Nazmul et~al.}(2005)\citenamefont{Nazmul, Amemiya,
  Shuto, Sugahara, and Tanaka}}]{Nazmul2005:PRL}
\bibinfo{author}{\bibfnamefont{A.~M.} \bibnamefont{Nazmul}},
  \bibinfo{author}{\bibfnamefont{T.}~\bibnamefont{Amemiya}},
  \bibinfo{author}{\bibfnamefont{Y.}~\bibnamefont{Shuto}},
  \bibinfo{author}{\bibfnamefont{S.}~\bibnamefont{Sugahara}}, \bibnamefont{and}
  \bibinfo{author}{\bibfnamefont{M.}~\bibnamefont{Tanaka}},
  \bibinfo{journal}{Phys. Rev. Lett.} \textbf{\bibinfo{volume}{95}},
  \bibinfo{pages}{017201} (\bibinfo{year}{2005}).

\bibitem[{\citenamefont{Saito et~al.}(2003)\citenamefont{Saito, Zayets,
  Yamagata, and Ando}}]{Saito2003:PRL}
\bibinfo{author}{\bibfnamefont{H.}~\bibnamefont{Saito}},
  \bibinfo{author}{\bibfnamefont{V.}~\bibnamefont{Zayets}},
  \bibinfo{author}{\bibfnamefont{S.}~\bibnamefont{Yamagata}}, \bibnamefont{and}
  \bibinfo{author}{\bibfnamefont{K.}~\bibnamefont{Ando}},
  \bibinfo{journal}{Phys. Rev. Lett.} \textbf{\bibinfo{volume}{90}},
  \bibinfo{pages}{207202} (\bibinfo{year}{2003}).

\bibitem[{\citenamefont{Pearton et~al.}(2003)\citenamefont{Pearton, Abernathy,
  Overberg, Thaler, Norton, Theodoropoulou, Hebard, Park, Ren, Kim
  et~al.}}]{Pearton2003:JAP}
\bibinfo{author}{\bibfnamefont{S.~J.} \bibnamefont{Pearton}},
  \bibinfo{author}{\bibfnamefont{C.~R.} \bibnamefont{Abernathy}},
  \bibinfo{author}{\bibfnamefont{M.~E.} \bibnamefont{Overberg}},
  \bibinfo{author}{\bibfnamefont{G.~T.} \bibnamefont{Thaler}},
  \bibinfo{author}{\bibfnamefont{D.~P.} \bibnamefont{Norton}},
  \bibinfo{author}{\bibfnamefont{N.}~\bibnamefont{Theodoropoulou}},
  \bibinfo{author}{\bibfnamefont{A.~F.} \bibnamefont{Hebard}},
  \bibinfo{author}{\bibfnamefont{Y.~D.} \bibnamefont{Park}},
  \bibinfo{author}{\bibfnamefont{F.}~\bibnamefont{Ren}},
  \bibinfo{author}{\bibfnamefont{J.}~\bibnamefont{Kim}}, \bibnamefont{et~al.},
  \bibinfo{journal}{J. Appl. Phys.} \textbf{\bibinfo{volume}{93}},
  \bibinfo{pages}{1} (\bibinfo{year}{2003}).

\bibitem[{\citenamefont{Erwin and {\v{Z}uti\'c}}(2004)}]{Erwin2004:NM}
\bibinfo{author}{\bibfnamefont{S.~C.} \bibnamefont{Erwin}} \bibnamefont{and}
  \bibinfo{author}{\bibfnamefont{I.}~\bibnamefont{{\v{Z}uti\'c}}},
  \bibinfo{journal}{Nature Mater.} \textbf{\bibinfo{volume}{3}},
  \bibinfo{pages}{410} (\bibinfo{year}{2004}).

\bibitem[{\citenamefont{Rashba}(2000)}]{Rashba2000:PRB}
\bibinfo{author}{\bibfnamefont{E.~I.} \bibnamefont{Rashba}},
  \bibinfo{journal}{Phys. Rev. B} \textbf{\bibinfo{volume}{62}},
  \bibinfo{pages}{R16267} (\bibinfo{year}{2000}).

\bibitem[{\citenamefont{Smith and Silver}(2001)}]{Smith2001:PRB}
\bibinfo{author}{\bibfnamefont{D.~L.} \bibnamefont{Smith}} \bibnamefont{and}
  \bibinfo{author}{\bibfnamefont{R.~N.} \bibnamefont{Silver}},
  \bibinfo{journal}{Phys. Rev. B} \textbf{\bibinfo{volume}{64}},
  \bibinfo{pages}{045323} (\bibinfo{year}{2001}).

\bibitem[{\citenamefont{Fert and Jaffres}(2001)}]{Fert2001:PRB}
\bibinfo{author}{\bibfnamefont{A.}~\bibnamefont{Fert}} \bibnamefont{and}
  \bibinfo{author}{\bibfnamefont{H.}~\bibnamefont{Jaffres}},
  \bibinfo{journal}{Phys. Rev. B} \textbf{\bibinfo{volume}{64}},
  \bibinfo{pages}{184420} (\bibinfo{year}{2001}).

\bibitem[{\citenamefont{Zega et~al.}(2006)\citenamefont{Zega, Hanbicki, Erwin,
  \v{Z}uti\'{c}, Kioseoglou, Li, Jonker, and Stroud}}]{Zega2006:PRL}
\bibinfo{author}{\bibfnamefont{T.~J.} \bibnamefont{Zega}},
  \bibinfo{author}{\bibfnamefont{A.~T.} \bibnamefont{Hanbicki}},
  \bibinfo{author}{\bibfnamefont{S.~C.} \bibnamefont{Erwin}},
  \bibinfo{author}{\bibfnamefont{I.}~\bibnamefont{\v{Z}uti\'{c}}},
  \bibinfo{author}{\bibfnamefont{G.}~\bibnamefont{Kioseoglou}},
  \bibinfo{author}{\bibfnamefont{C.~H.} \bibnamefont{Li}},
  \bibinfo{author}{\bibfnamefont{B.~T.} \bibnamefont{Jonker}},
  \bibnamefont{and} \bibinfo{author}{\bibfnamefont{R.~M.}
  \bibnamefont{Stroud}}, \bibinfo{journal}{Phys. Rev. Lett}
  \textbf{\bibinfo{volume}{96}}, \bibinfo{pages}{196101}
  (\bibinfo{year}{2006}).

\bibitem[{\citenamefont{Parkin et~al.}(2004)\citenamefont{Parkin, Kaiser,
  Panchula, Rice, Samant, and Yang}}]{Parkin2004:P}
\bibinfo{author}{\bibfnamefont{S.~S.~P.} \bibnamefont{Parkin}},
  \bibinfo{author}{\bibfnamefont{C.}~\bibnamefont{Kaiser}},
  \bibinfo{author}{\bibfnamefont{A.}~\bibnamefont{Panchula}},
  \bibinfo{author}{\bibfnamefont{P.}~\bibnamefont{Rice}},
  \bibinfo{author}{\bibfnamefont{M.}~\bibnamefont{Samant}}, \bibnamefont{and}
  \bibinfo{author}{\bibfnamefont{S.-H.} \bibnamefont{Yang}},
  \bibinfo{journal}{{\sl Nature Mater.}} \textbf{\bibinfo{volume}{3}},
  \bibinfo{pages}{862} (\bibinfo{year}{2004}).

\bibitem[{\citenamefont{Yuasa et~al.}(2004)\citenamefont{Yuasa, Nagahama,
  Fukushira, Suzuki, and Ando}}]{Yuasa2004:NM}
\bibinfo{author}{\bibfnamefont{S.}~\bibnamefont{Yuasa}},
  \bibinfo{author}{\bibfnamefont{T.}~\bibnamefont{Nagahama}},
  \bibinfo{author}{\bibfnamefont{A.}~\bibnamefont{Fukushira}},
  \bibinfo{author}{\bibfnamefont{Y.}~\bibnamefont{Suzuki}}, \bibnamefont{and}
  \bibinfo{author}{\bibfnamefont{K.}~\bibnamefont{Ando}},
  \bibinfo{journal}{{\sl Nature Mater.}} \textbf{\bibinfo{volume}{3}},
  \bibinfo{pages}{868} (\bibinfo{year}{2004}).

\bibitem[{\citenamefont{Butler et~al.}(2001)\citenamefont{Butler, Zhang,
  Schulthes, and {MacLaren}}}]{Butler2001:PRB}
\bibinfo{author}{\bibfnamefont{W.~H.} \bibnamefont{Butler}},
  \bibinfo{author}{\bibfnamefont{X.}~\bibnamefont{Zhang}},
  \bibinfo{author}{\bibfnamefont{T.~C.} \bibnamefont{Schulthes}},
  \bibnamefont{and} \bibinfo{author}{\bibfnamefont{J.~M.}
  \bibnamefont{{MacLaren}}}, \bibinfo{journal}{Phys. Rev. B}
  \textbf{\bibinfo{volume}{63}}, \bibinfo{pages}{054416}
  (\bibinfo{year}{2001}).

\bibitem[{\citenamefont{Mathon and Umerski}(2001)}]{Mathon2001:PRB}
\bibinfo{author}{\bibfnamefont{J.}~\bibnamefont{Mathon}} \bibnamefont{and}
  \bibinfo{author}{\bibfnamefont{A.}~\bibnamefont{Umerski}},
  \bibinfo{journal}{Phys. Rev. B} \textbf{\bibinfo{volume}{63}},
  \bibinfo{pages}{220403} (\bibinfo{year}{2001}).

\bibitem[{\citenamefont{Jiang et~al.}(2005)\citenamefont{Jiang, Wang, Shelby,
  Macfarlane, Bank, Harris, and Parkin}}]{Jiang2005:PRL}
\bibinfo{author}{\bibfnamefont{X.}~\bibnamefont{Jiang}},
  \bibinfo{author}{\bibfnamefont{R.}~\bibnamefont{Wang}},
  \bibinfo{author}{\bibfnamefont{R.~M.} \bibnamefont{Shelby}},
  \bibinfo{author}{\bibfnamefont{R.~M.} \bibnamefont{Macfarlane}},
  \bibinfo{author}{\bibfnamefont{S.~R.} \bibnamefont{Bank}},
  \bibinfo{author}{\bibfnamefont{J.~S.} \bibnamefont{Harris}},
  \bibnamefont{and} \bibinfo{author}{\bibfnamefont{S.~S.~P.}
  \bibnamefont{Parkin}}, \bibinfo{journal}{Phys. Rev. Lett.}
  \textbf{\bibinfo{volume}{94}}, \bibinfo{pages}{056601}
  (\bibinfo{year}{2005}).

\bibitem[{\citenamefont{Wang et~al.}(2005)\citenamefont{Wang, Jiang, Shelby,
  Macfarlane, Parkin, Bank, and Harris}}]{Wang2005:APL}
\bibinfo{author}{\bibfnamefont{R.}~\bibnamefont{Wang}},
  \bibinfo{author}{\bibfnamefont{X.}~\bibnamefont{Jiang}},
  \bibinfo{author}{\bibfnamefont{R.~M.} \bibnamefont{Shelby}},
  \bibinfo{author}{\bibfnamefont{R.~M.} \bibnamefont{Macfarlane}},
  \bibinfo{author}{\bibfnamefont{S.~S.~P.} \bibnamefont{Parkin}},
  \bibinfo{author}{\bibfnamefont{S.~R.} \bibnamefont{Bank}}, \bibnamefont{and}
  \bibinfo{author}{\bibfnamefont{J.~S.} \bibnamefont{Harris}},
  \bibinfo{journal}{Appl. Phys. Lett.} \textbf{\bibinfo{volume}{86}},
  \bibinfo{pages}{052901} (\bibinfo{year}{2005}).

\bibitem[{\citenamefont{Salis et~al.}(2005)\citenamefont{Salis, Wang, Jiang,
  Shelby, Parkin, Bank, and Harris}}]{Salis2005:APL}
\bibinfo{author}{\bibfnamefont{G.}~\bibnamefont{Salis}},
  \bibinfo{author}{\bibfnamefont{R.}~\bibnamefont{Wang}},
  \bibinfo{author}{\bibfnamefont{X.}~\bibnamefont{Jiang}},
  \bibinfo{author}{\bibfnamefont{R.~M.} \bibnamefont{Shelby}},
  \bibinfo{author}{\bibfnamefont{S.~S.~P.} \bibnamefont{Parkin}},
  \bibinfo{author}{\bibfnamefont{S.~R.} \bibnamefont{Bank}}, \bibnamefont{and}
  \bibinfo{author}{\bibfnamefont{J.~S.} \bibnamefont{Harris}},
  \bibinfo{journal}{Appl. Phys. Lett.} \textbf{\bibinfo{volume}{87}},
  \bibinfo{pages}{262503} (\bibinfo{year}{2005}).

\bibitem[{\citenamefont{Mott}(1936)}]{Mott1936:PRCa}
\bibinfo{author}{\bibfnamefont{N.~F.} \bibnamefont{Mott}},
  \bibinfo{journal}{Proc. R. Soc. London, Ser. A}
  \textbf{\bibinfo{volume}{153}}, \bibinfo{pages}{699} (\bibinfo{year}{1936}).

\bibitem[{\citenamefont{Jonker et~al.}(2004)\citenamefont{Jonker, Hanbicki,
  Pierece, and Stiles}}]{Jonker2003:P}
\bibinfo{author}{\bibfnamefont{B.~T.} \bibnamefont{Jonker}},
  \bibinfo{author}{\bibfnamefont{A.~T.} \bibnamefont{Hanbicki}},
  \bibinfo{author}{\bibfnamefont{D.~T.} \bibnamefont{Pierece}},
  \bibnamefont{and} \bibinfo{author}{\bibfnamefont{M.~D.}
  \bibnamefont{Stiles}}, \bibinfo{journal}{J. Magn. Magn. Mater.}
  \textbf{\bibinfo{volume}{277}}, \bibinfo{pages}{24} (\bibinfo{year}{2004}).

\bibitem[{\citenamefont{Gijs and Bauer}(1997)}]{Gijs1997:AP}
\bibinfo{author}{\bibfnamefont{M.~A.~M.} \bibnamefont{Gijs}} \bibnamefont{and}
  \bibinfo{author}{\bibfnamefont{G.~E.~W.} \bibnamefont{Bauer}},
  \bibinfo{journal}{Adv. Phys.} \textbf{\bibinfo{volume}{46}},
  \bibinfo{pages}{285} (\bibinfo{year}{1997}).

\bibitem[{\citenamefont{Johnson and Silsbee}(1988)}]{Johnson1988:PRBa}
\bibinfo{author}{\bibfnamefont{M.}~\bibnamefont{Johnson}} \bibnamefont{and}
  \bibinfo{author}{\bibfnamefont{R.~H.} \bibnamefont{Silsbee}},
  \bibinfo{journal}{Phys. Rev. B} \textbf{\bibinfo{volume}{37}},
  \bibinfo{pages}{5312} (\bibinfo{year}{1988}).

\bibitem[{\citenamefont{Stiles and Zangwill}(2002)}]{Stiles2002:PRB}
\bibinfo{author}{\bibfnamefont{M.~D.} \bibnamefont{Stiles}} \bibnamefont{and}
  \bibinfo{author}{\bibfnamefont{A.}~\bibnamefont{Zangwill}},
  \bibinfo{journal}{Phys. Rev. B} \textbf{\bibinfo{volume}{66}},
  \bibinfo{pages}{014407} (\bibinfo{year}{2002}).

\bibitem[{\citenamefont{Hershfield and Zhao}(1997)}]{Hershfield1997:PRB}
\bibinfo{author}{\bibfnamefont{S.}~\bibnamefont{Hershfield}} \bibnamefont{and}
  \bibinfo{author}{\bibfnamefont{H.~L.} \bibnamefont{Zhao}},
  \bibinfo{journal}{Phys. Rev. B} \textbf{\bibinfo{volume}{56}},
  \bibinfo{pages}{3296} (\bibinfo{year}{1997}).

\bibitem[{\citenamefont{Rashba}(2002)}]{Rashba2002:EPJ}
\bibinfo{author}{\bibfnamefont{E.~I.} \bibnamefont{Rashba}},
  \bibinfo{journal}{Eur. Phys. J. B} \textbf{\bibinfo{volume}{29}},
  \bibinfo{pages}{513} (\bibinfo{year}{2002}).

\bibitem[{\citenamefont{Takahashi and Maekawa}(2003)}]{Takahashi2003:PRB}
\bibinfo{author}{\bibfnamefont{S.}~\bibnamefont{Takahashi}} \bibnamefont{and}
  \bibinfo{author}{\bibfnamefont{S.}~\bibnamefont{Maekawa}},
  \bibinfo{journal}{Phys. Rev. B} \textbf{\bibinfo{volume}{67}},
  \bibinfo{pages}{052409} (\bibinfo{year}{2003}).

\bibitem[{\citenamefont{Jonker et~al.}(2003)\citenamefont{Jonker, Erwin,
  Petrou, and Petukhov}}]{Jonker2003:MRS}
\bibinfo{author}{\bibfnamefont{B.~T.} \bibnamefont{Jonker}},
  \bibinfo{author}{\bibfnamefont{S.~C.} \bibnamefont{Erwin}},
  \bibinfo{author}{\bibfnamefont{A.}~\bibnamefont{Petrou}}, \bibnamefont{and}
  \bibinfo{author}{\bibfnamefont{A.~G.} \bibnamefont{Petukhov}},
  \bibinfo{journal}{MRS Bull.} \textbf{\bibinfo{volume}{28}},
  \bibinfo{pages}{740} (\bibinfo{year}{2003}).

\bibitem[{\citenamefont{Jedema et~al.}(2001)\citenamefont{Jedema, Filip, and
  {van Wees}}}]{Jedema2001:N}
\bibinfo{author}{\bibfnamefont{F.~J.} \bibnamefont{Jedema}},
  \bibinfo{author}{\bibfnamefont{A.~T.} \bibnamefont{Filip}}, \bibnamefont{and}
  \bibinfo{author}{\bibfnamefont{B.~J.} \bibnamefont{{van Wees}}},
  \bibinfo{journal}{{\sl Nature}} \textbf{\bibinfo{volume}{410}},
  \bibinfo{pages}{345} (\bibinfo{year}{2001}).

\bibitem[{\citenamefont{Valet and Fert}(1993)}]{Valet1993:PRB}
\bibinfo{author}{\bibfnamefont{T.}~\bibnamefont{Valet}} \bibnamefont{and}
  \bibinfo{author}{\bibfnamefont{A.}~\bibnamefont{Fert}},
  \bibinfo{journal}{Phys. Rev. B} \textbf{\bibinfo{volume}{48}},
  \bibinfo{pages}{7099} (\bibinfo{year}{1993}).

\bibitem[{\citenamefont{Johnson and Silsbee}(1985)}]{Johnson1985:PRL}
\bibinfo{author}{\bibfnamefont{M.}~\bibnamefont{Johnson}} \bibnamefont{and}
  \bibinfo{author}{\bibfnamefont{R.~H.} \bibnamefont{Silsbee}},
  \bibinfo{journal}{Phys. Rev. Lett.} \textbf{\bibinfo{volume}{55}},
  \bibinfo{pages}{1790} (\bibinfo{year}{1985}).

\bibitem[{\citenamefont{Garzon et~al.}(2005)\citenamefont{Garzon, \v{Z}uti\'c,
  and Webb}}]{Garzon2005:PRL}
\bibinfo{author}{\bibfnamefont{S.}~\bibnamefont{Garzon}},
  \bibinfo{author}{\bibfnamefont{I.}~\bibnamefont{\v{Z}uti\'c}},
  \bibnamefont{and} \bibinfo{author}{\bibfnamefont{R.~A.} \bibnamefont{Webb}},
  \bibinfo{journal}{Phys. Rev. Lett} \textbf{\bibinfo{volume}{94}},
  \bibinfo{pages}{176601} (\bibinfo{year}{2005}).

\bibitem[{\citenamefont{Godfrey and Johnson}(2006)}]{Godfrey2006:PRL}
\bibinfo{author}{\bibfnamefont{R.}~\bibnamefont{Godfrey}} \bibnamefont{and}
  \bibinfo{author}{\bibfnamefont{M.}~\bibnamefont{Johnson}},
  \bibinfo{journal}{Phys. Rev. Lett.} \textbf{\bibinfo{volume}{95}},
  \bibinfo{pages}{136601} (\bibinfo{year}{2006}).

\bibitem[{\citenamefont{{\v{Z}uti\'{c}}
  et~al.}(2001{\natexlab{a}})\citenamefont{{\v{Z}uti\'{c}}, Fabian, and {Das
  Sarma}}}]{Zutic2001:PRB}
\bibinfo{author}{\bibfnamefont{I.}~\bibnamefont{{\v{Z}uti\'{c}}}},
  \bibinfo{author}{\bibfnamefont{J.}~\bibnamefont{Fabian}}, \bibnamefont{and}
  \bibinfo{author}{\bibfnamefont{S.}~\bibnamefont{{Das Sarma}}},
  \bibinfo{journal}{Phys. Rev. B} \textbf{\bibinfo{volume}{64}},
  \bibinfo{pages}{121201} (\bibinfo{year}{2001}{\natexlab{a}}).

\bibitem[{\citenamefont{{\v{Z}uti\'{c}}
  et~al.}(2002)\citenamefont{{\v{Z}uti\'{c}}, Fabian, and {Das
  Sarma}}}]{Zutic2002:PRL}
\bibinfo{author}{\bibfnamefont{I.}~\bibnamefont{{\v{Z}uti\'{c}}}},
  \bibinfo{author}{\bibfnamefont{J.}~\bibnamefont{Fabian}}, \bibnamefont{and}
  \bibinfo{author}{\bibfnamefont{S.}~\bibnamefont{{Das Sarma}}},
  \bibinfo{journal}{Phys. Rev. Lett.} \textbf{\bibinfo{volume}{88}},
  \bibinfo{pages}{066603} (\bibinfo{year}{2002}).

\bibitem[{\citenamefont{Fabian et~al.}(2002{\natexlab{a}})\citenamefont{Fabian,
  {\v{Z}uti\'{c}}, and {Das Sarma}}}]{Fabian2002:PRB}
\bibinfo{author}{\bibfnamefont{J.}~\bibnamefont{Fabian}},
  \bibinfo{author}{\bibfnamefont{I.}~\bibnamefont{{\v{Z}uti\'{c}}}},
  \bibnamefont{and} \bibinfo{author}{\bibfnamefont{S.}~\bibnamefont{{Das
  Sarma}}}, \bibinfo{journal}{Phys. Rev. B} \textbf{\bibinfo{volume}{66}},
  \bibinfo{pages}{165301} (\bibinfo{year}{2002}{\natexlab{a}}).

\bibitem[{\citenamefont{Meier and {Zakharchenya (Eds.)}}(1984)}]{Meier:1984}
\bibinfo{author}{\bibfnamefont{F.}~\bibnamefont{Meier}} \bibnamefont{and}
  \bibinfo{author}{\bibfnamefont{B.~P.} \bibnamefont{{Zakharchenya (Eds.)}}},
  \emph{\bibinfo{title}{Optical Orientation}}
  (\bibinfo{publisher}{North-Holand, New York}, \bibinfo{year}{1984}).

\bibitem[{\citenamefont{{\v{Z}uti\'{c}}
  et~al.}(2006{\natexlab{a}})\citenamefont{{\v{Z}uti\'{c}}, Fabian, and
  Erwin}}]{Zutic2004:P}
\bibinfo{author}{\bibfnamefont{I.}~\bibnamefont{{\v{Z}uti\'{c}}}},
  \bibinfo{author}{\bibfnamefont{J.}~\bibnamefont{Fabian}}, \bibnamefont{and}
  \bibinfo{author}{\bibfnamefont{S.~C.} \bibnamefont{Erwin}},
  \bibinfo{journal}{Phys. Rev. Lett.} \textbf{\bibinfo{volume}{97}},
  \bibinfo{pages}{026602} (\bibinfo{year}{2006}{\natexlab{a}}).

\bibitem[{\citenamefont{{\v{Z}uti\'{c}}
  et~al.}(2001{\natexlab{b}})\citenamefont{{\v{Z}uti\'{c}}, Fabian, and {Das
  Sarma}}}]{Zutic2001:APL}
\bibinfo{author}{\bibfnamefont{I.}~\bibnamefont{{\v{Z}uti\'{c}}}},
  \bibinfo{author}{\bibfnamefont{J.}~\bibnamefont{Fabian}}, \bibnamefont{and}
  \bibinfo{author}{\bibfnamefont{S.}~\bibnamefont{{Das Sarma}}},
  \bibinfo{journal}{Appl. Phys. Lett.} \textbf{\bibinfo{volume}{79}},
  \bibinfo{pages}{1558} (\bibinfo{year}{2001}{\natexlab{b}}).

\bibitem[{\citenamefont{{\v{Z}uti\'{c}}
  et~al.}(2003)\citenamefont{{\v{Z}uti\'{c}}, Fabian, and {Das
  Sarma}}}]{Zutic2003:APL}
\bibinfo{author}{\bibfnamefont{I.}~\bibnamefont{{\v{Z}uti\'{c}}}},
  \bibinfo{author}{\bibfnamefont{J.}~\bibnamefont{Fabian}}, \bibnamefont{and}
  \bibinfo{author}{\bibfnamefont{S.}~\bibnamefont{{Das Sarma}}},
  \bibinfo{journal}{Appl. Phys. Lett.} \textbf{\bibinfo{volume}{82}},
  \bibinfo{pages}{221} (\bibinfo{year}{2003}).

\bibitem[{\citenamefont{Ashcroft and Mermin}(1976)}]{Ashcroft:1976}
\bibinfo{author}{\bibfnamefont{N.~W.} \bibnamefont{Ashcroft}} \bibnamefont{and}
  \bibinfo{author}{\bibfnamefont{N.~D.} \bibnamefont{Mermin}},
  \emph{\bibinfo{title}{Solid State Physics}} (\bibinfo{publisher}{Sounders,
  Philadelphia}, \bibinfo{year}{1976}).

\bibitem[{\citenamefont{Tse et~al.}(2005)\citenamefont{Tse, Fabian,
  \v{Z}uti\'{c}, and {Das Sarma}}}]{Tse2005:PRB}
\bibinfo{author}{\bibfnamefont{W.-K.} \bibnamefont{Tse}},
  \bibinfo{author}{\bibfnamefont{J.}~\bibnamefont{Fabian}},
  \bibinfo{author}{\bibfnamefont{I.}~\bibnamefont{\v{Z}uti\'{c}}},
  \bibnamefont{and} \bibinfo{author}{\bibfnamefont{S.}~\bibnamefont{{Das
  Sarma}}}, \bibinfo{journal}{Phys. Rev. B} \textbf{\bibinfo{volume}{72}},
  \bibinfo{pages}{214303} (\bibinfo{year}{2005}).

\bibitem[{\citenamefont{Weisbuch}(1977)}]{Weisbuch1977:T}
\bibinfo{author}{\bibfnamefont{C.}~\bibnamefont{Weisbuch}},
  \bibinfo{journal}{Ph.D. Thesis (Paris Univ. VIII) p. 26}
  (\bibinfo{year}{1977}).

\bibitem[{\citenamefont{Dzhioev et~al.}(1997)\citenamefont{Dzhioev,
  Zakharchenya, Korenev, and Stepanova}}]{Dzhioev1997:PSS}
\bibinfo{author}{\bibfnamefont{R.~I.} \bibnamefont{Dzhioev}},
  \bibinfo{author}{\bibfnamefont{B.~P.} \bibnamefont{Zakharchenya}},
  \bibinfo{author}{\bibfnamefont{V.~L.} \bibnamefont{Korenev}},
  \bibnamefont{and} \bibinfo{author}{\bibfnamefont{M.~N.}
  \bibnamefont{Stepanova}}, \bibinfo{journal}{Fiz. Tverd. Tela}
  \textbf{\bibinfo{volume}{39}}, \bibinfo{pages}{1975} (\bibinfo{year}{1997}),
  \bibinfo{note}{[Phys. Solid State {\bf 39}, 1765-1768 (1997)]}.

\bibitem[{\citenamefont{Kikkawa and Awschalom}(1998)}]{Kikkawa1998:PRL}
\bibinfo{author}{\bibfnamefont{J.~M.} \bibnamefont{Kikkawa}} \bibnamefont{and}
  \bibinfo{author}{\bibfnamefont{D.~D.} \bibnamefont{Awschalom}},
  \bibinfo{journal}{Phys. Rev. Lett.} \textbf{\bibinfo{volume}{80}},
  \bibinfo{pages}{4313} (\bibinfo{year}{1998}).

\bibitem[{\citenamefont{Dzhioev et~al.}(2001)\citenamefont{Dzhioev,
  Zakharchenya, Korenev, Gammon, and Katzer}}]{Dzhioev2001:JETPL}
\bibinfo{author}{\bibfnamefont{R.~I.} \bibnamefont{Dzhioev}},
  \bibinfo{author}{\bibfnamefont{B.~P.} \bibnamefont{Zakharchenya}},
  \bibinfo{author}{\bibfnamefont{V.~L.} \bibnamefont{Korenev}},
  \bibinfo{author}{\bibfnamefont{D.}~\bibnamefont{Gammon}}, \bibnamefont{and}
  \bibinfo{author}{\bibfnamefont{D.~S.} \bibnamefont{Katzer}},
  \bibinfo{journal}{Zh. Eksp. Teor. Fiz. Pisma Red.}
  \textbf{\bibinfo{volume}{74}}, \bibinfo{pages}{182} (\bibinfo{year}{2001}),
  \bibinfo{note}{[JETP Lett. {\bf 74}, 200-203 (2001)]}.

\bibitem[{\citenamefont{Hilton and Tang}(2002)}]{Hilton2002:PRL}
\bibinfo{author}{\bibfnamefont{D.~J.} \bibnamefont{Hilton}} \bibnamefont{and}
  \bibinfo{author}{\bibfnamefont{C.~L.} \bibnamefont{Tang}},
  \bibinfo{journal}{Phys. Rev. Lett.} \textbf{\bibinfo{volume}{89}},
  \bibinfo{pages}{146601} (\bibinfo{year}{2002}).

\bibitem[{\citenamefont{{\v{Z}uti\'{c}}
  et~al.}(2006{\natexlab{b}})\citenamefont{{\v{Z}uti\'{c}}, Fabian, and
  Erwin}}]{Zutic2006:IBM}
\bibinfo{author}{\bibfnamefont{I.}~\bibnamefont{{\v{Z}uti\'{c}}}},
  \bibinfo{author}{\bibfnamefont{J.}~\bibnamefont{Fabian}}, \bibnamefont{and}
  \bibinfo{author}{\bibfnamefont{S.~C.} \bibnamefont{Erwin}},
  \bibinfo{journal}{IBM J. Res. \& Dev.} \textbf{\bibinfo{volume}{50}},
  \bibinfo{pages}{121} (\bibinfo{year}{2006}{\natexlab{b}}).

\bibitem[{\citenamefont{Ganichev and Prettl}(2003)}]{Ganichev2003:JPCM}
\bibinfo{author}{\bibfnamefont{S.~D.} \bibnamefont{Ganichev}} \bibnamefont{and}
  \bibinfo{author}{\bibfnamefont{W.}~\bibnamefont{Prettl}},
  \bibinfo{journal}{J. Phys.: Condens. Matter.} \textbf{\bibinfo{volume}{15}},
  \bibinfo{pages}{R935} (\bibinfo{year}{2003}).

\bibitem[{\citenamefont{Long et~al.}(2003)\citenamefont{Long, Sun, Guo, and
  Wang}}]{Long2002:APL}
\bibinfo{author}{\bibfnamefont{W.}~\bibnamefont{Long}},
  \bibinfo{author}{\bibfnamefont{Q.-F.} \bibnamefont{Sun}},
  \bibinfo{author}{\bibfnamefont{H.}~\bibnamefont{Guo}}, \bibnamefont{and}
  \bibinfo{author}{\bibfnamefont{J.}~\bibnamefont{Wang}},
  \bibinfo{journal}{Appl. Phys. Lett.} \textbf{\bibinfo{volume}{83}},
  \bibinfo{pages}{1397} (\bibinfo{year}{2003}).

\bibitem[{\citenamefont{Mal'shukov et~al.}(2003)\citenamefont{Mal'shukov, Tang,
  Chu, and Chao}}]{Malshukov2003:PRB}
\bibinfo{author}{\bibfnamefont{A.~G.} \bibnamefont{Mal'shukov}},
  \bibinfo{author}{\bibfnamefont{C.~S.} \bibnamefont{Tang}},
  \bibinfo{author}{\bibfnamefont{C.~S.} \bibnamefont{Chu}}, \bibnamefont{and}
  \bibinfo{author}{\bibfnamefont{K.~A.} \bibnamefont{Chao}},
  \bibinfo{journal}{Phys. Rev. B} \textbf{\bibinfo{volume}{68}},
  \bibinfo{pages}{233307} (\bibinfo{year}{2003}).

\bibitem[{\citenamefont{Datta}(2005)}]{Datta2005:APL}
\bibinfo{author}{\bibfnamefont{S.}~\bibnamefont{Datta}},
  \bibinfo{journal}{Appl. Phys. Lett.} \textbf{\bibinfo{volume}{87}},
  \bibinfo{pages}{013115} (\bibinfo{year}{2005}).

\bibitem[{\citenamefont{Pershin and Privman}(2003)}]{Pershin2003:PRL}
\bibinfo{author}{\bibfnamefont{Y.~V.} \bibnamefont{Pershin}} \bibnamefont{and}
  \bibinfo{author}{\bibfnamefont{V.}~\bibnamefont{Privman}},
  \bibinfo{journal}{Phys. Rev. Lett.} \textbf{\bibinfo{volume}{90}},
  \bibinfo{pages}{256602} (\bibinfo{year}{2003}).

\bibitem[{\citenamefont{Furdyna}(1988)}]{Furdyna1988:JAP}
\bibinfo{author}{\bibfnamefont{J.~K.} \bibnamefont{Furdyna}},
  \bibinfo{journal}{J. Appl. Phys.} \textbf{\bibinfo{volume}{64}},
  \bibinfo{pages}{R29} (\bibinfo{year}{1988}).

\bibitem[{\citenamefont{Dietl}(1994)}]{Dietl:1994}
\bibinfo{author}{\bibfnamefont{T.}~\bibnamefont{Dietl}}, in
  \emph{\bibinfo{booktitle}{Handbook of Semiconductors, {Vol. 3}}}, edited by
  \bibinfo{editor}{\bibfnamefont{T.~S.} \bibnamefont{Moss}} \bibnamefont{and}
  \bibinfo{editor}{\bibfnamefont{S.}~\bibnamefont{Mahajan}}
  (\bibinfo{publisher}{{North-Holland}, New York}, \bibinfo{year}{1994}), p.
  \bibinfo{pages}{1251}.

\bibitem[{\citenamefont{Zudov et~al.}(2002)\citenamefont{Zudov, Kono, Matsuda,
  Ikaida, Miura, Munekata, Sanders, Sun, and Stanton}}]{Zudov2002:PRB}
\bibinfo{author}{\bibfnamefont{M.~A.} \bibnamefont{Zudov}},
  \bibinfo{author}{\bibfnamefont{J.}~\bibnamefont{Kono}},
  \bibinfo{author}{\bibfnamefont{Y.~H.} \bibnamefont{Matsuda}},
  \bibinfo{author}{\bibfnamefont{T.}~\bibnamefont{Ikaida}},
  \bibinfo{author}{\bibfnamefont{N.}~\bibnamefont{Miura}},
  \bibinfo{author}{\bibfnamefont{H.}~\bibnamefont{Munekata}},
  \bibinfo{author}{\bibfnamefont{G.~D.} \bibnamefont{Sanders}},
  \bibinfo{author}{\bibfnamefont{Y.}~\bibnamefont{Sun}}, \bibnamefont{and}
  \bibinfo{author}{\bibfnamefont{C.~J.} \bibnamefont{Stanton}},
  \bibinfo{journal}{Phys. Rev. B} \textbf{\bibinfo{volume}{66}},
  \bibinfo{pages}{161307} (\bibinfo{year}{2002}).

\bibitem[{\citenamefont{Deutsch et~al.}(2004)\citenamefont{Deutsch, Vignale,
  and Flatt\'{e}}}]{Deutsch2004:JAP}
\bibinfo{author}{\bibfnamefont{M.}~\bibnamefont{Deutsch}},
  \bibinfo{author}{\bibfnamefont{G.}~\bibnamefont{Vignale}}, \bibnamefont{and}
  \bibinfo{author}{\bibfnamefont{M.~F.} \bibnamefont{Flatt\'{e}}},
  \bibinfo{journal}{J. Appl. Phys.} \textbf{\bibinfo{volume}{96}},
  \bibinfo{pages}{7424} (\bibinfo{year}{2004}).

\bibitem[{\citenamefont{Schmeltzer et~al.}(2003)\citenamefont{Schmeltzer,
  Saxena, Bishop, and Smith}}]{Schmeltzer2003:PRB}
\bibinfo{author}{\bibfnamefont{D.}~\bibnamefont{Schmeltzer}},
  \bibinfo{author}{\bibfnamefont{A.}~\bibnamefont{Saxena}},
  \bibinfo{author}{\bibfnamefont{A.}~\bibnamefont{Bishop}}, \bibnamefont{and}
  \bibinfo{author}{\bibfnamefont{D.~L.} \bibnamefont{Smith}},
  \bibinfo{journal}{Phys. Rev. B} \textbf{\bibinfo{volume}{68}},
  \bibinfo{pages}{195317} (\bibinfo{year}{2003}).

\bibitem[{\citenamefont{Schmidt and Molenkamp}(2002)}]{Schmidt2002:SST}
\bibinfo{author}{\bibfnamefont{G.}~\bibnamefont{Schmidt}} \bibnamefont{and}
  \bibinfo{author}{\bibfnamefont{L.~W.} \bibnamefont{Molenkamp}},
  \bibinfo{journal}{Semicond. Sci. Technol.} \textbf{\bibinfo{volume}{17}},
  \bibinfo{pages}{310} (\bibinfo{year}{2002}).

\bibitem[{\citenamefont{Stephens et~al.}(2004)\citenamefont{Stephens,
  Berezovsky, McGuire, Sham, Gossard, and Awschalom}}]{Stephens2004:PRL}
\bibinfo{author}{\bibfnamefont{J.}~\bibnamefont{Stephens}},
  \bibinfo{author}{\bibfnamefont{J.}~\bibnamefont{Berezovsky}},
  \bibinfo{author}{\bibfnamefont{J.~P.} \bibnamefont{McGuire}},
  \bibinfo{author}{\bibfnamefont{L.~J.} \bibnamefont{Sham}},
  \bibinfo{author}{\bibfnamefont{A.~C.} \bibnamefont{Gossard}},
  \bibnamefont{and} \bibinfo{author}{\bibfnamefont{D.~D.}
  \bibnamefont{Awschalom}}, \bibinfo{journal}{Phys. Rev. Lett.}
  \textbf{\bibinfo{volume}{93}}, \bibinfo{pages}{097602}
  (\bibinfo{year}{2004}).

\bibitem[{\citenamefont{Bratkovsky and Osipov}(2004)}]{Bratkovsky2004:JAP}
\bibinfo{author}{\bibfnamefont{A.~M.} \bibnamefont{Bratkovsky}}
  \bibnamefont{and} \bibinfo{author}{\bibfnamefont{V.~V.}
  \bibnamefont{Osipov}}, \bibinfo{journal}{J. Appl. Phys.}
  \textbf{\bibinfo{volume}{96}}, \bibinfo{pages}{4525} (\bibinfo{year}{2004}).

\bibitem[{\citenamefont{Osipov et~al.}(2005)\citenamefont{Osipov, Petukhov, and
  Smelyankiy}}]{Osipov2005:APL}
\bibinfo{author}{\bibfnamefont{V.~V.} \bibnamefont{Osipov}},
  \bibinfo{author}{\bibfnamefont{A.~G.} \bibnamefont{Petukhov}},
  \bibnamefont{and} \bibinfo{author}{\bibfnamefont{V.~N.}
  \bibnamefont{Smelyankiy}}, \bibinfo{journal}{Appl. Phys. Lett.}
  \textbf{\bibinfo{volume}{87}}, \bibinfo{pages}{202112}
  (\bibinfo{year}{2005}).

\bibitem[{\citenamefont{Petukhov et~al.}(2006)\citenamefont{Petukhov,
  Smelyankiy, and Osipov}}]{Petukhov2006:P}
\bibinfo{author}{\bibfnamefont{A.~G.} \bibnamefont{Petukhov}},
  \bibinfo{author}{\bibfnamefont{V.~N.} \bibnamefont{Smelyankiy}},
  \bibnamefont{and} \bibinfo{author}{\bibfnamefont{V.~V.} \bibnamefont{Osipov}}
  (\bibinfo{year}{2006}), \eprint{cond-mat/0609599}.

\bibitem[{\citenamefont{Silsbee}(1980)}]{Silsbee1980:BMR}
\bibinfo{author}{\bibfnamefont{R.~H.} \bibnamefont{Silsbee}},
  \bibinfo{journal}{Bull. Magn. Reson.} \textbf{\bibinfo{volume}{2}},
  \bibinfo{pages}{284} (\bibinfo{year}{1980}).

\bibitem[{\citenamefont{{\v{Z}uti\'{c}} and Fabian}(2003)}]{Zutic2003:P}
\bibinfo{author}{\bibfnamefont{I.}~\bibnamefont{{\v{Z}uti\'{c}}}}
  \bibnamefont{and} \bibinfo{author}{\bibfnamefont{J.}~\bibnamefont{Fabian}},
  \bibinfo{journal}{Mater. Trans., JIM} \textbf{\bibinfo{volume}{44}},
  \bibinfo{pages}{2062} (\bibinfo{year}{2003}).

\bibitem[{\citenamefont{Zhao et~al.}(2002)\citenamefont{Zhao, Matsukara, Abe,
  Chiba, Ohno, Takamura, and Ohno}}]{Zhao2002:JCG}
\bibinfo{author}{\bibfnamefont{J.~H.} \bibnamefont{Zhao}},
  \bibinfo{author}{\bibfnamefont{F.}~\bibnamefont{Matsukara}},
  \bibinfo{author}{\bibfnamefont{E.}~\bibnamefont{Abe}},
  \bibinfo{author}{\bibfnamefont{D.}~\bibnamefont{Chiba}},
  \bibinfo{author}{\bibfnamefont{Y.}~\bibnamefont{Ohno}},
  \bibinfo{author}{\bibfnamefont{K.}~\bibnamefont{Takamura}}, \bibnamefont{and}
  \bibinfo{author}{\bibfnamefont{H.}~\bibnamefont{Ohno}}, \bibinfo{journal}{J.
  Cryst. Growth} \textbf{\bibinfo{volume}{237-239}}, \bibinfo{pages}{1349}
  (\bibinfo{year}{2002}).

\bibitem[{\citenamefont{Ishida et~al.}(2003)\citenamefont{Ishida, Sarma,
  Okazaki, Hwang, Ott, Fujimori, Medvedkin, Ishibashi, and
  Sato}}]{Ishida2003:PRL}
\bibinfo{author}{\bibfnamefont{Y.}~\bibnamefont{Ishida}},
  \bibinfo{author}{\bibfnamefont{D.~D.} \bibnamefont{Sarma}},
  \bibinfo{author}{\bibfnamefont{K.}~\bibnamefont{Okazaki}},
  \bibinfo{author}{\bibfnamefont{J.~O. J.~I.} \bibnamefont{Hwang}},
  \bibinfo{author}{\bibfnamefont{H.}~\bibnamefont{Ott}},
  \bibinfo{author}{\bibfnamefont{A.}~\bibnamefont{Fujimori}},
  \bibinfo{author}{\bibfnamefont{G.~A.} \bibnamefont{Medvedkin}},
  \bibinfo{author}{\bibfnamefont{T.}~\bibnamefont{Ishibashi}},
  \bibnamefont{and} \bibinfo{author}{\bibfnamefont{K.}~\bibnamefont{Sato}},
  \bibinfo{journal}{Phys. Rev. Lett.} \textbf{\bibinfo{volume}{91}},
  \bibinfo{pages}{107202} (\bibinfo{year}{2003}).

\bibitem[{\citenamefont{Kondo et~al.}(2006)\citenamefont{Kondo, Hayafuji, and
  Munekata}}]{Kondo2006:JJAP}
\bibinfo{author}{\bibfnamefont{T.}~\bibnamefont{Kondo}},
  \bibinfo{author}{\bibfnamefont{J.}~\bibnamefont{Hayafuji}}, \bibnamefont{and}
  \bibinfo{author}{\bibfnamefont{H.}~\bibnamefont{Munekata}},
  \bibinfo{journal}{Jpn. J. Appl. Phys.} \textbf{\bibinfo{volume}{45}},
  \bibinfo{pages}{L663} (\bibinfo{year}{2006}).

\bibitem[{\citenamefont{Chen et~al.}(2006)\citenamefont{Chen, Moser, Kotissek,
  Sadowski, Zenger, Weiss, and Wegscheider}}]{Chen2006:P}
\bibinfo{author}{\bibfnamefont{P.}~\bibnamefont{Chen}},
  \bibinfo{author}{\bibfnamefont{J.}~\bibnamefont{Moser}},
  \bibinfo{author}{\bibfnamefont{P.}~\bibnamefont{Kotissek}},
  \bibinfo{author}{\bibfnamefont{J.}~\bibnamefont{Sadowski}},
  \bibinfo{author}{\bibfnamefont{M.}~\bibnamefont{Zenger}},
  \bibinfo{author}{\bibfnamefont{D.}~\bibnamefont{Weiss}}, \bibnamefont{and}
  \bibinfo{author}{\bibfnamefont{W.}~\bibnamefont{Wegscheider}}
  (\bibinfo{year}{2006}), \eprint{cond-mat/0608453}.

\bibitem[{\citenamefont{Nakagawa et~al.}(2005)\citenamefont{Nakagawa, Asai,
  Mukunoki, Susaki, and Hwang}}]{Nakagawa2004:P}
\bibinfo{author}{\bibfnamefont{N.}~\bibnamefont{Nakagawa}},
  \bibinfo{author}{\bibfnamefont{M.}~\bibnamefont{Asai}},
  \bibinfo{author}{\bibfnamefont{Y.}~\bibnamefont{Mukunoki}},
  \bibinfo{author}{\bibfnamefont{T.}~\bibnamefont{Susaki}}, \bibnamefont{and}
  \bibinfo{author}{\bibfnamefont{H.~Y.} \bibnamefont{Hwang}},
  \bibinfo{journal}{Appl. Phys. Lett.} \textbf{\bibinfo{volume}{86}},
  \bibinfo{pages}{082504} (\bibinfo{year}{2005}).

\bibitem[{\citenamefont{Min et~al.}(2006)\citenamefont{Min, Motohashi, Lodder,
  and Jansen}}]{Min2006:NM}
\bibinfo{author}{\bibfnamefont{B.-C.} \bibnamefont{Min}},
  \bibinfo{author}{\bibfnamefont{K.}~\bibnamefont{Motohashi}},
  \bibinfo{author}{\bibfnamefont{C.}~\bibnamefont{Lodder}}, \bibnamefont{and}
  \bibinfo{author}{\bibfnamefont{R.}~\bibnamefont{Jansen}},
  \bibinfo{journal}{Nature Mater.} \textbf{\bibinfo{volume}{5}},
  \bibinfo{pages}{817} (\bibinfo{year}{2006}).

\bibitem[{\citenamefont{{\v{Z}uti\'{c}}}(2006)}]{Zutic2006:NM}
\bibinfo{author}{\bibfnamefont{I.}~\bibnamefont{{\v{Z}uti\'{c}}}},
  \bibinfo{journal}{Nature Mater.} \textbf{\bibinfo{volume}{5}},
  \bibinfo{pages}{771} (\bibinfo{year}{2006}).

\bibitem[{\citenamefont{Dennis et~al.}(2003)\citenamefont{Dennis, Tiusan,
  Ferreira, Gregg, Ensell, Thompson, and Freitas}}]{Dennis2005:JMMM}
\bibinfo{author}{\bibfnamefont{C.~L.} \bibnamefont{Dennis}},
  \bibinfo{author}{\bibfnamefont{C.~V.} \bibnamefont{Tiusan}},
  \bibinfo{author}{\bibfnamefont{R.~A.} \bibnamefont{Ferreira}},
  \bibinfo{author}{\bibfnamefont{J.~F.} \bibnamefont{Gregg}},
  \bibinfo{author}{\bibfnamefont{G.~J.} \bibnamefont{Ensell}},
  \bibinfo{author}{\bibfnamefont{S.~M.} \bibnamefont{Thompson}},
  \bibnamefont{and} \bibinfo{author}{\bibfnamefont{P.~P.}
  \bibnamefont{Freitas}}, \bibinfo{journal}{J. Magn. Magn. Mater.}
  \textbf{\bibinfo{volume}{1383}}, \bibinfo{pages}{290} (\bibinfo{year}{2003}).

\bibitem[{\citenamefont{Jansen}(2003)}]{Jansen2003:JPD}
\bibinfo{author}{\bibfnamefont{R.}~\bibnamefont{Jansen}}, \bibinfo{journal}{J.
  Phys. D. Appl. Phys.} \textbf{\bibinfo{volume}{36}}, \bibinfo{pages}{R289}
  (\bibinfo{year}{2003}).

\bibitem[{\citenamefont{Mizushima et~al.}(1997)\citenamefont{Mizushima, Kinno,
  Yamauchi, and Tanaka}}]{Mizushima1997:IEEETM}
\bibinfo{author}{\bibfnamefont{K.}~\bibnamefont{Mizushima}},
  \bibinfo{author}{\bibfnamefont{T.}~\bibnamefont{Kinno}},
  \bibinfo{author}{\bibfnamefont{T.}~\bibnamefont{Yamauchi}}, \bibnamefont{and}
  \bibinfo{author}{\bibfnamefont{K.}~\bibnamefont{Tanaka}},
  \bibinfo{journal}{IEEE Trans. Magn.} \textbf{\bibinfo{volume}{33}},
  \bibinfo{pages}{3500} (\bibinfo{year}{1997}).

\bibitem[{\citenamefont{Yamauchi and Mizushima}(1998)}]{Yamauchi1998:PRB}
\bibinfo{author}{\bibfnamefont{T.}~\bibnamefont{Yamauchi}} \bibnamefont{and}
  \bibinfo{author}{\bibfnamefont{K.}~\bibnamefont{Mizushima}},
  \bibinfo{journal}{Phys. Rev. B} \textbf{\bibinfo{volume}{58}},
  \bibinfo{pages}{1934} (\bibinfo{year}{1998}).

\bibitem[{\citenamefont{Sato and Mizushima}(2001)}]{Sato2001:APL}
\bibinfo{author}{\bibfnamefont{R.}~\bibnamefont{Sato}} \bibnamefont{and}
  \bibinfo{author}{\bibfnamefont{K.}~\bibnamefont{Mizushima}},
  \bibinfo{journal}{Appl. Phys. Lett.} \textbf{\bibinfo{volume}{79}},
  \bibinfo{pages}{1157} (\bibinfo{year}{2001}).

\bibitem[{\citenamefont{{van Dijken} et~al.}(2002)\citenamefont{{van Dijken},
  Jiang, and Parkin}}]{vanDijken2002:APL}
\bibinfo{author}{\bibfnamefont{S.}~\bibnamefont{{van Dijken}}},
  \bibinfo{author}{\bibfnamefont{X.}~\bibnamefont{Jiang}}, \bibnamefont{and}
  \bibinfo{author}{\bibfnamefont{S.~S.~P.} \bibnamefont{Parkin}},
  \bibinfo{journal}{Appl. Phys. Lett.} \textbf{\bibinfo{volume}{80}},
  \bibinfo{pages}{3364} (\bibinfo{year}{2002}).

\bibitem[{\citenamefont{{van Dijken}
  et~al.}(2003{\natexlab{a}})\citenamefont{{van Dijken}, Jiang, and
  Parkin}}]{vanDijken2003:APLa}
\bibinfo{author}{\bibfnamefont{S.}~\bibnamefont{{van Dijken}}},
  \bibinfo{author}{\bibfnamefont{X.}~\bibnamefont{Jiang}}, \bibnamefont{and}
  \bibinfo{author}{\bibfnamefont{S.~S.~P.} \bibnamefont{Parkin}},
  \bibinfo{journal}{Appl. Phys. Lett.} \textbf{\bibinfo{volume}{82}},
  \bibinfo{pages}{775} (\bibinfo{year}{2003}{\natexlab{a}}).

\bibitem[{\citenamefont{{van Dijken}
  et~al.}(2003{\natexlab{b}})\citenamefont{{van Dijken}, Jiang, and
  Parkin}}]{vanDijken2003:PRL}
\bibinfo{author}{\bibfnamefont{S.}~\bibnamefont{{van Dijken}}},
  \bibinfo{author}{\bibfnamefont{X.}~\bibnamefont{Jiang}}, \bibnamefont{and}
  \bibinfo{author}{\bibfnamefont{S.~S.~P.} \bibnamefont{Parkin}},
  \bibinfo{journal}{Phys. Rev. Lett.} \textbf{\bibinfo{volume}{90}},
  \bibinfo{pages}{197203} (\bibinfo{year}{2003}{\natexlab{b}}).

\bibitem[{\citenamefont{Huang et~al.}(2004)\citenamefont{Huang, Lo, Yao, Hsieh,
  Ju, Huang, and Huang}}]{Huang2004:APL}
\bibinfo{author}{\bibfnamefont{Y.~W.} \bibnamefont{Huang}},
  \bibinfo{author}{\bibfnamefont{C.~K.} \bibnamefont{Lo}},
  \bibinfo{author}{\bibfnamefont{Y.~D.} \bibnamefont{Yao}},
  \bibinfo{author}{\bibfnamefont{L.~C.} \bibnamefont{Hsieh}},
  \bibinfo{author}{\bibfnamefont{J.~J.} \bibnamefont{Ju}},
  \bibinfo{author}{\bibfnamefont{D.~R.} \bibnamefont{Huang}}, \bibnamefont{and}
  \bibinfo{author}{\bibfnamefont{J.~H.} \bibnamefont{Huang}},
  \bibinfo{journal}{Appl. Phys. Lett.} \textbf{\bibinfo{volume}{85}},
  \bibinfo{pages}{2959} (\bibinfo{year}{2004}).

\bibitem[{\citenamefont{Huang et~al.}(2005)\citenamefont{Huang, Lo, Yao, Hsieh,
  and Huang}}]{Huang2005:JAP}
\bibinfo{author}{\bibfnamefont{Y.~W.} \bibnamefont{Huang}},
  \bibinfo{author}{\bibfnamefont{C.~K.} \bibnamefont{Lo}},
  \bibinfo{author}{\bibfnamefont{Y.~D.} \bibnamefont{Yao}},
  \bibinfo{author}{\bibfnamefont{L.~C.} \bibnamefont{Hsieh}}, \bibnamefont{and}
  \bibinfo{author}{\bibfnamefont{J.~H.} \bibnamefont{Huang}},
  \bibinfo{journal}{J. Appl. Phys.} \textbf{\bibinfo{volume}{97}},
  \bibinfo{pages}{10D504} (\bibinfo{year}{2005}).

\bibitem[{\citenamefont{Fabian et~al.}(2002{\natexlab{b}})\citenamefont{Fabian,
  \v{Z}uti\'{c}, and {Das Sarma}}}]{Fabian2002:P}
\bibinfo{author}{\bibfnamefont{J.}~\bibnamefont{Fabian}},
  \bibinfo{author}{\bibfnamefont{I.}~\bibnamefont{\v{Z}uti\'{c}}},
  \bibnamefont{and} \bibinfo{author}{\bibfnamefont{S.}~\bibnamefont{{Das
  Sarma}}} (\bibinfo{year}{2002}{\natexlab{b}}), \eprint{cond-mat/0211639}.

\bibitem[{\citenamefont{Flatt{\'{e}} et~al.}(2003)\citenamefont{Flatt{\'{e}},
  Yu, Johnston-Halperin, and Awschalom}}]{Flatte2003:APL}
\bibinfo{author}{\bibfnamefont{M.~E.} \bibnamefont{Flatt{\'{e}}}},
  \bibinfo{author}{\bibfnamefont{Z.~G.} \bibnamefont{Yu}},
  \bibinfo{author}{\bibfnamefont{E.}~\bibnamefont{Johnston-Halperin}},
  \bibnamefont{and} \bibinfo{author}{\bibfnamefont{D.~D.}
  \bibnamefont{Awschalom}}, \bibinfo{journal}{Appl. Phys. Lett.}
  \textbf{\bibinfo{volume}{82}}, \bibinfo{pages}{4740} (\bibinfo{year}{2003}).

\bibitem[{\citenamefont{Lebedeva and Kuivalainen}(2003)}]{Lebedeva2003:JAP}
\bibinfo{author}{\bibfnamefont{N.}~\bibnamefont{Lebedeva}} \bibnamefont{and}
  \bibinfo{author}{\bibfnamefont{P.}~\bibnamefont{Kuivalainen}},
  \bibinfo{journal}{J. Appl. Phys.} \textbf{\bibinfo{volume}{93}},
  \bibinfo{pages}{9845} (\bibinfo{year}{2003}).

\bibitem[{\citenamefont{Dery et~al.}(2006)\citenamefont{Dery, Cywinski, and
  Sham}}]{Dery2006:PRB}
\bibinfo{author}{\bibfnamefont{H.}~\bibnamefont{Dery}},
  \bibinfo{author}{\bibfnamefont{L.}~\bibnamefont{Cywinski}}, \bibnamefont{and}
  \bibinfo{author}{\bibfnamefont{L.~J.} \bibnamefont{Sham}},
  \bibinfo{journal}{Phys. Rev. B} \textbf{\bibinfo{volume}{73}},
  \bibinfo{pages}{161307} (\bibinfo{year}{2006}).

\bibitem[{\citenamefont{Fabian and \v{Z}uti\'{c}}(2005)}]{Fabian2005:APL}
\bibinfo{author}{\bibfnamefont{J.}~\bibnamefont{Fabian}} \bibnamefont{and}
  \bibinfo{author}{\bibfnamefont{I.}~\bibnamefont{\v{Z}uti\'{c}}},
  \bibinfo{journal}{Appl. Phys. Lett.} \textbf{\bibinfo{volume}{86}},
  \bibinfo{pages}{133506} (\bibinfo{year}{2005}).

\bibitem[{\citenamefont{Fabian et~al.}(2004)\citenamefont{Fabian,
  \v{Z}uti\'{c}, and Sarma}}]{Fabian2004:APL}
\bibinfo{author}{\bibfnamefont{J.}~\bibnamefont{Fabian}},
  \bibinfo{author}{\bibfnamefont{I.}~\bibnamefont{\v{Z}uti\'{c}}},
  \bibnamefont{and} \bibinfo{author}{\bibfnamefont{S.~D.} \bibnamefont{Sarma}},
  \bibinfo{journal}{Appl. Phys. Lett.} \textbf{\bibinfo{volume}{84}},
  \bibinfo{pages}{85} (\bibinfo{year}{2004}).

\bibitem[{\citenamefont{Fabian and {\v{Z}uti\'{c}}}(2004)}]{Fabian2004:PRB}
\bibinfo{author}{\bibfnamefont{J.}~\bibnamefont{Fabian}} \bibnamefont{and}
  \bibinfo{author}{\bibfnamefont{I.}~\bibnamefont{{\v{Z}uti\'{c}}}},
  \bibinfo{journal}{Phys. Rev. B} \textbf{\bibinfo{volume}{69}},
  \bibinfo{pages}{115314} (\bibinfo{year}{2004}).

\bibitem[{\citenamefont{Fabian and \v{Z}uti\'{c}}(2004)}]{Fabian2004:APP}
\bibinfo{author}{\bibfnamefont{J.}~\bibnamefont{Fabian}} \bibnamefont{and}
  \bibinfo{author}{\bibfnamefont{I.}~\bibnamefont{\v{Z}uti\'{c}}},
  \bibinfo{journal}{Acta Physica Polonica A} \textbf{\bibinfo{volume}{106}},
  \bibinfo{pages}{109} (\bibinfo{year}{2004}).

\bibitem[{\citenamefont{Tiwari}(1992)}]{Tiwari:1992}
\bibinfo{author}{\bibfnamefont{S.}~\bibnamefont{Tiwari}},
  \emph{\bibinfo{title}{Compound Semiconductor Device Physics}}
  (\bibinfo{publisher}{Academic Press, San Diego}, \bibinfo{year}{1992}).

\end{thebibliography}
\end{document}